\documentclass{aa}
\usepackage{graphicx}
\usepackage[usenames]{color} 

\definecolor{xblue}{rgb}{0,0,1}

\begin{document}

\title{
Weak magnetic fields in central stars of planetary nebulae?\thanks{Based on
observations obtained at the European Southern Observatory,
Paranal, Chile (ESO program No.~088.D-0425(A)).}
}

   \author{M.~Steffen\inst{1}
          \and
          S.~Hubrig\inst{1}
          \and
          H.~Todt\inst{2}
          \and
          M.~Sch\"oller\inst{3}
          \and
          W.-R.~Hamann\inst{2} 
          \and 
          C.~Sandin\inst{1}
          \and 
          D.~Sch\"onberner\inst{1} 
}

   \institute{Leibniz-Institut f\"ur Astrophysik Potsdam,
             An der Sternwarte~16,
             14482~Potsdam, Germany
         \and
             Universit\"at Potsdam, Institut f\"ur Physik und Astronomie, 
             14476~Potsdam, Germany
         \and
             European Southern Observatory,
              Karl-Schwarzschild-Str.~2,
              85748 Garching, Germany
             }

   \date{Received March 19, 2014; accepted August 13, 2014}

 
  \abstract
   {
It is not yet clear whether magnetic fields play an essential role in shaping
  planetary nebulae (PNe), or whether stellar rotation alone and/or a close
  binary companion, stellar or substellar, can account for the variety of the 
  observed nebular morphologies.
   }
   {
In a quest for empirical evidence verifying or disproving the role of 
magnetic fields in shaping planetary nebulae, we follow up on previous attempts 
to measure the magnetic field in a representative sample of PN central stars. 
   }
   {
We obtained low-resolution polarimetric spectra with FORS\,2 installed on 
the Antu telescope of the VLT for a sample of 12 bright 
central stars of PNe with different morphologies, including two round nebulae,
seven elliptical nebulae, and three bipolar nebulae. Two targets are Wolf-Rayet
type central stars.
   }
   {
For the majority of the observed central stars, we do not find any
significant evidence for the existence of surface magnetic fields. However,
our measurements may indicate the presence of weak mean longitudinal magnetic 
fields of the order of $100$ Gauss in the central star of the young
elliptical planetary nebula IC\,418 as well as in the Wolf-Rayet type
central star of the bipolar nebula Hen\,2-113 and the weak emission line 
central star of the elliptical nebula Hen\,2-131.
A clear detection of a $250$~G mean longitudinal field is achieved for
the A-type companion of the central star of NGC\,1514. Some of the 
central stars show a moderate night-to-night spectrum
variability, which may be the signature of a variable stellar wind  
and/or rotational modulation due to magnetic features.
   }
   {
Since our analysis indicates only weak fields, if any, in a few targets of our
sample, we conclude that strong magnetic fields of the order of kG are not
widespread among PNe central stars. Nevertheless, simple estimates based on
a theoretical model of magnetized wind bubbles suggest that even weak
magnetic fields below the current detection limit of the order of 100~G may
well be sufficient to contribute to the shaping of the surrounding nebulae
throughout their evolution. Our current sample is too small to draw
conclusions about a correlation between nebular morphology and the presence of
stellar magnetic fields.
   }


   \maketitle

\section{Introduction}
\label{sect:intro}
One of the major open questions regarding the formation of planetary nebulae
(PNe) concerns the mechanism that is responsible for their non-spherical, 
often axisymmetric shaping (e.g., review by Balick \& Frank 
\cite{BalickFrank2002}). Both central star binarity and
stellar rotation in combination with magnetic fields are among the favorite
explanations, but their role in PN formation and shaping is not yet 
sufficiently clear.

Basically, the origin of PNe is understood 
to be a consequence of the interaction of the hot central star with 
its circumstellar environment through photoionization and wind-wind collision.
In the framework of this theory, the formation and evolution of PNe has
been modeled in detail, assuming spherical symmetry
(e.g., Marten \& Sch\"onberner \cite{MartenSchoenberner1991}; 
Mellema \cite{Mellema1994}; Villaver at al.\ \cite{Villaver2002};
Perinotto et al.\ \cite{Perinotto2004}; 
Steffen \& Sch\"onberner \cite{Steffen2006}) or axisymmetric geometry 
(e.g., Mellema \cite{Mellema1995,Mellema1997}). The role of stellar rotation 
and magnetic fields in shaping PNe has been studied numerically in the 
pioneering work of Garc{\'{\i}}a-Segura et al.\ (\cite{GarciaSegura1999}), 
who find that most of the observed nebular morphologies can be modeled
with an appropriate combination of input parameters. Magnetic shaping of PNe
thus appears to be an attractive alternative to the popular binary 
hypothesis (e.g., De Marco \cite{DeMarco2009}; Douchin et al.\ 
\cite{Douchin2012}).
It remains unclear, however, whether the underlying assumptions of 
 Garc{\'{\i}}a-Segura et al.\ (constant stellar wind with a high magnetization
parameter) are actually realistic, since very little is known so far about 
rotation rates and surface magnetic fields of the central stars of PNe.

The discovery of sufficiently strong magnetic fields in the central stars 
would lend considerable support to the hypothesis that the axisymmetric
or bipolar appearance of many PNe is caused by
magnetic fields (e.g., Garc{\'{\i}}a-D{\'{\i}}az et al.\ \cite{GarciaDiaz2008};
Blackman \cite{Blackman2009}). On the other hand,
theoretical arguments have been put forward to demonstrate that the 
structure of non-spherical planetary nebulae cannot be attributed 
to the presence of large scale magnetic fields, since they would
contain more angular momentum and energy than a single star can 
supply (Soker \cite{Soker2006}). 

\begin{table*}[htb]
\caption{
List of PNe central stars observed in the framework of our program,
ordered by increasing RA.
}
\label{tab:objects}
\centering
\begin{tabular}{llrccclrr}
\hline
\hline\noalign{\smallskip}
\multicolumn{1}{c}{(1)} &
\multicolumn{1}{c}{(2)} &
\multicolumn{1}{c}{(3)} &
\multicolumn{1}{c}{(4)} &
\multicolumn{1}{c}{(5)} &
\multicolumn{1}{c}{(6)} &
\multicolumn{1}{l}{(7)} \\
\multicolumn{1}{c}{Name}      &
\multicolumn{1}{c}{Catalog}   &
\multicolumn{1}{c}{$m_{\rm V}$}    &
\multicolumn{2}{c}{Spectral type} &
\multicolumn{1}{c}{Nebular}   &
\multicolumn{1}{l}{X-ray$^{b)}$}     &
\multicolumn{2}{c}{References} \\
 &
\multicolumn{1}{c}{Identifier} &
 &
\multicolumn{1}{c}{CSPN}       &
\multicolumn{1}{c}{Companion$^{a)}$}  &
\multicolumn{1}{c}{morphology} &
\multicolumn{1}{l}{emission}   &
\multicolumn{1}{c}{col. (4)}   &
\multicolumn{1}{c}{col. (6)}  \\
\hline\noalign{\smallskip}
\object{NGC\,246}             & 118.8-74.7  & 11.96  & PG1159  & G8-K0V & R  & SP    & $^{1)}$    &  $^{6)}$  \\
\object{NGC\,1360}$^{\,\ast)}$  & 220.3-53.9  & 11.35  & sdO     &        & E  & SP   & $^{2)}$     &   $^{6)}$  \\
\object{NGC\,1514}            & 165.5-15.2  & 9.42   &      -- & A0III  & E  & HP    &            &  $^{6)}$  \\
\object{IC\,418}              & 215.2-24.2  & 10.17  & O7f     &        & E  & D     & $^{3)}$     &  $^{6)}$  \\
\object{NGC\,2346}            & 215.6+03.6  & 11.47  &      -- & A5V    & B  & N     &            &  $^{7)}$  \\
\object{NGC\,2392}            & 197.8+17.3  & 10.53  & O6f     & dM?    & E  & HP, D & $^{3)}$     &  $^{6)}$  \\
\object{Hen\,2-36}            & 279.6-03.1  & 11.30  &      -- & A2III  & B  & F     &            &  $^{7)}$  \\
\object{LSS\,1362}$^{\,\ast)}$  & 273.6+06.1  & 12.47  & sdO     &        & E  &      & $^{4)}$      &  $^{8)}$  \\
\object{NGC\,3132}            & 272.1+12.3  & 10.07  &      -- & A2IV-V & E  & N     &             &  $^{6)}$  \\
\object{Hen\,2-108}           & 316.1+08.4  & 12.72  &  WELS   &        & R  &       & $^{5)}$      &  $^{6)}$  \\
\object{Hen\,2-113}           & 321.0+03.9  & 12.28  & [WC11]  &        & B  & F     & $^{2),\,5)}$ &  $^{9)}$  \\
\object{Hen\,2-131}           & 315.1-13.0  & 11.01  &  WELS   &        & E  & F     & $^{2),\,5)}$ &  $^{6)}$  \\
\hline\noalign{\smallskip}
\end{tabular}
\tablefoot{
(2): Identification number from the Strasbourg Catalog of PNe; 
(3): Integrated $V$-band intensity of central star; \\
(4): Spectral classification of central star ([WC11]: Wolf-Rayet type; 
     WELS: weak emission line star, normal chemical composition); \\
(5): Spectral classification of  binary companion (A-type companions 
     dominate the spectrum, spectral type of CSPN unknown); \\ 
(6): Basic morphology of the planetary nebula, R (round), E (elliptical), 
     B (bipolar); 
(7): X-ray classification, N (non-detection), SP (soft point source), 
     HP (hard point source), D (diffuse source), F (future Chandra target); \\
$^{\,\ast)}$target analyzed by Jordan et al.\ (\cite{Jordan2005,Jordan2012});
$^{a)}$De Marco (\cite{DeMarco2009}); 
$^{b)}$Kastner et al.\ (\cite{Kastner2012}); \\
$^{1)}$Werner \& Herwig (\cite{WernerHerwig2006}); 
$^{2)}$Mendez \& Niemela (\cite{MendezNiemela1977}); 
$^{3)}$Heap (\cite{Heap1977}); 
$^{4)}$Drilling (\cite{Drilling1983}); 
$^{5)}$Tylenda et al.\ (\cite{Tylenda1993}); \\
$^{6)}$Phillips (\cite{Phillips2003}); 
$^{7)}$Corradi \& Schwarz (\cite{Corradi1995}); 
$^{8)}$Jordan et al.\ (\cite{Jordan2005}); 
$^{9)}$Lagadec et al.\ (\cite{Lagadec2006}). }
\end{table*}

In principle, the role of magnetic fields in shaping PNe  may be verified or 
disproved by empirical evidence, as already suggested by Jordan et 
al.\ (\cite{Jordan2005}).
Using FORS\,1 in spectropolarimetric mode, they reported the detection of 
magnetic fields of the order of kG in the central stars of 
the PNe NGC\,1360 and LSS\,1362. 
A reanalysis of their data, however, did not provide any significant evidence 
for longitudinal magnetic fields in these stars that are stronger than a few 
hundred Gauss (Jordan et al.\ \cite{Jordan2012}).
Their field measurements have typical error bars of 150 to 300\,G.
Similar results were achieved in the work of Leone et 
al.\ (\cite{Leone2011}), who concluded that the mean longitudinal magnetic 
fields in NGC\,1360 and LSS\,1362 are much weaker,
less than 600\,G, or that the field has a complex structure.
The most recent search for magnetic fields in central stars of planetary 
nebulae by Leone et al.\ (\cite{Leone2014}), based on spectropolarimetric 
observations of 19 central stars with WHT/ISIS and VLT/FORS\,2, is partly
affected by large measurement uncertainties and reports
no positive detection either.
Thus, convincing evidence for the presence of significant magnetic fields in 
the central stars of PNe is still missing. 

Using low-resolution polarimetric spectra obtained with FORS\,2 installed at
the Very Large Telescope, we carried out a search for magnetic fields in a
sample of 12 central stars covering the whole range of morphologies from 
round to elliptical/axisymmetric, and bipolar PNe, and including both 
chemically normal and Wolf-Rayet (WR) type (hydrogen-poor) central stars. The 
sample includes two round nebulae (NGC\,246, Hen\,2-108), five elliptical 
nebulae (IC\,418, NGC\,1514, NGC\,2392, NGC\,3132, Hen\,2-131), and three
 bipolar nebulae (NGC\,2346, Hen\,2-36, Hen\,2-113). Two targets are 
WR-type central stars (NGC\,246, Hen\,2-113). In addition, we included 
the two (elliptical) targets of Jordan et al.\ (\cite{Jordan2005}), 
NGC\,1360 and LSS\,1362, for which they originally claimed the detection of 
kG magnetic fields. Six of the 12 central stars are known binaries.

The data collected in our survey can provide a basis for further empirical
investigations. Since our sample comprises both normal hydrogen-burning, and
Wolf-Rayet type (hydrogen-deficient) central stars,  some insight may be
gained into the physics of WR-type winds regarding the question of whether 
the much enhanced mass loss of [WC] type central stars (by a factor of
$\sim$100 with respect to the mass loss of normal central stars) is related to
the presence of stellar magnetic fields. 

Our target list includes a number of nebulae whose X-ray emission has been 
measured by Chandra and/or XMM Newton or are prospective targets in the 
Chandra PN Large Project (PI J.~Kastner). In principle, our magnetic field 
measurements may help to answer the question why some central stars show up 
as X-ray point sources, while others, with similar
stellar parameters, do not emit X-rays. The magnetic properties of the central
stars and their winds could play an important role in this
context. Similarly, the central cavity of some PNe is known to be a source of
diffuse thermal X-ray emission, while other PNe are undetected in diffuse
X-rays. Again, magnetic fields may be responsible: (i) sufficiently strong 
magnetic fields are expected to modify the thermal structure of the 
shock-heated stellar wind, and hence its X-ray emission, and (ii)
even very weak magnetic fields suppress thermal conduction perpendicular 
to the field lines, with severe consequences for the X-ray luminosity
and characteristic temperature (Steffen et al.\ \cite{Steffen2008}).

In Sect.\,\ref{sect:observations}, we give an overview of our observations 
and magnetic field measurements with VLT/FORS\,2, before we discuss the main
results of our 12 targets in Sect.\,\ref{sect:individual}. Section
\ref{sect:discussion} is devoted to theoretical estimates of the role of 
the central star's magnetic fields in PN shaping and of the influence of 
the magnetic fields on the diffuse X-ray emission of PNe. Finally, we
summarize our main conclusions in Sect.\,\ref{sect:conclusions}.

\section{Observations and magnetic field measurements}
\label{sect:observations}

\begin{figure*}
\centering
\includegraphics[angle=0,width=0.94\textwidth]{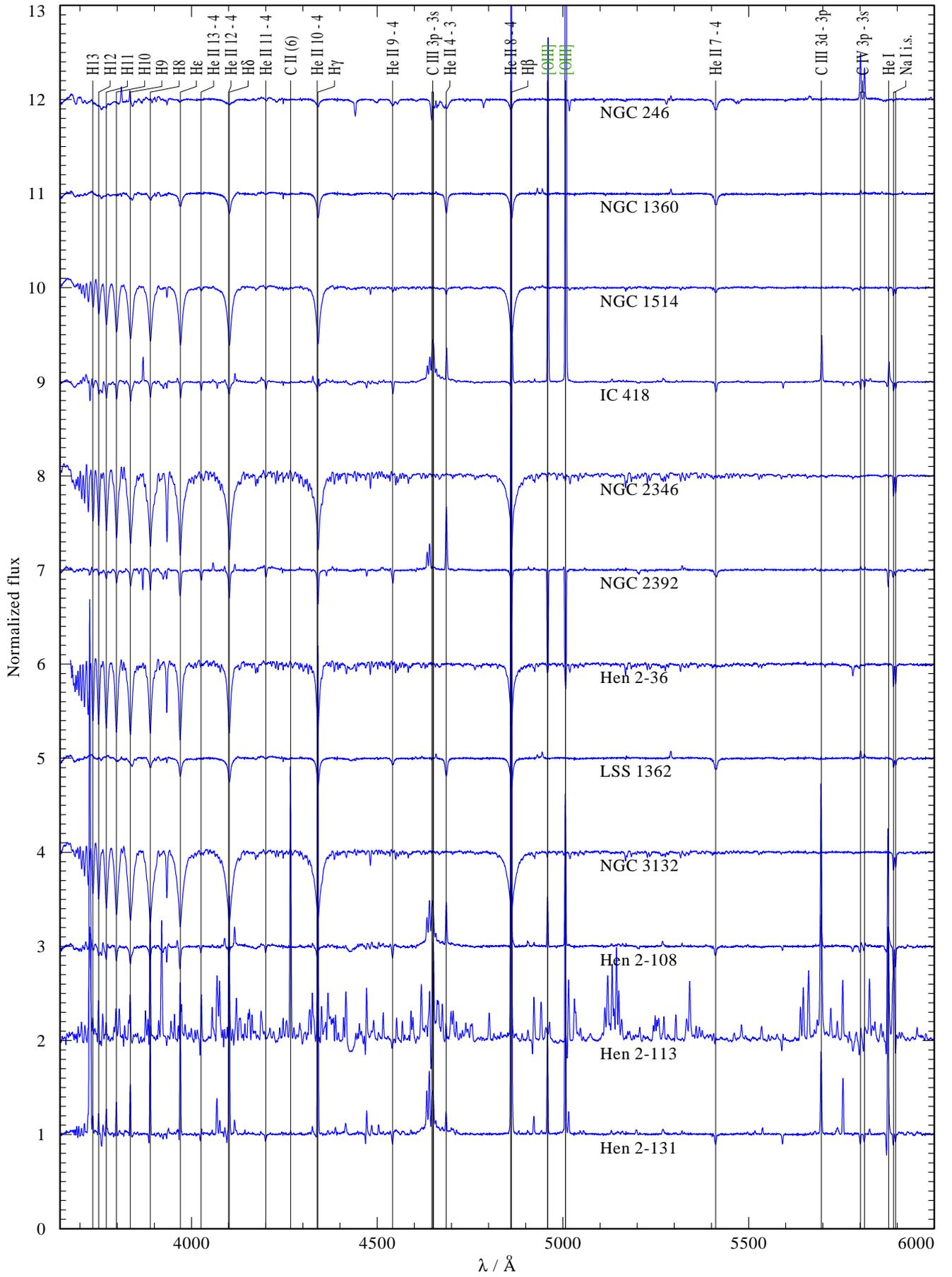}
\caption{
Normalized FORS\,2 Stokes~$I$ spectra of the $12$ central stars in our sample,
displayed with a vertical offset of 1 unit between adjacent spectra.
The strongest spectral lines are identified.
The forbidden [\ion{O}{iii}] nebular emission lines close to 5000\,\AA{} 
are marked in green. They may appear in absorption 
(e.g., NGC\,2392) due to imperfect background subtraction.}
\label{fig:norm}
\end{figure*}

Spectropolarimetric observations of 12 central stars were carried
out from 2011 October 5 to 2012 March 28 in service mode at the European
Southern Observatory with FORS\,2 mounted on the 8-m Antu telescope of the
VLT. Nine targets were observed twice, and three targets
were observed three times. We present the list of targets in our sample 
in Table~\ref{tab:objects}, where we also collect information about the
spectral classification of the central stars, binarity, morphology of the 
nebulae, and observed X-ray emission.  

The multi-mode instrument FORS\,2 is equipped with polarization analyzing
optics, comprising superachromatic halfwave and quarterwave phase retarder
plates, and a Wollaston prism with a beam divergence of 22$\arcsec$ in
standard resolution mode.  During the observations with FORS\,2, we used a
slit width of 0\farcs5 and the GRISM 600B to achieve a spectral resolving
power of about 1650. 

From the raw FORS\,2 data, the parallel and perpendicular beams
are extracted using a pipeline written in the MIDAS environment
by T.~Szeifert, the very first FORS instrument scientist.
This pipeline reduction by default includes background subtraction.
A unique wavelength calibration frame is used for each night.

A first description of the assessment of the longitudinal magnetic field
measurements using FORS\,1/2 spectropolarimetric observations was presented 
in our previous work (e.g., Hubrig et al.\ \cite{Hubrig2004a,Hubrig2004b}, 
and references therein).
To minimize the crosstalk effect, a sequence of subexposures at the retarder
position angles $-45^{\circ}+45^{\circ}$, $+45^{\circ}-45^{\circ}$,
$-45^{\circ}+45^{\circ}$, etc., is usually executed during observations and
the $V/I$ spectrum is calculated using:
\begin{equation}
\frac{V}{I} = \frac{1}{2} \left\{ 
\left( \frac{f^{\rm o} - f^{\rm e}}{f^{\rm o} + f^{\rm e}} \right)_{-45^{\circ}} -
\left( \frac{f^{\rm o} - f^{\rm e}}{f^{\rm o} + f^{\rm e}} \right)_{+45^{\circ}} \right\}
\end{equation}
where $+45^{\circ}$ and $-45^{\circ}$ indicate the position angle of the
retarder waveplate and $f^{\rm o}$ and $f^{\rm e}$ are the ordinary and
extraordinary beams, respectively.  Rectification of the $V/I$ spectra was
performed in the way described by Hubrig et al.\ (\cite{Hubrig2014}).
Null profiles, $N$, are calculated as pairwise differences from all available 
$V$ profiles.  From these, 3$\sigma$-outliers are identified and used to clip 
the $V$ profiles.  This removes spurious signals, which mostly come from cosmic
rays and also reduces the noise. A full description of the updated data 
reduction and analysis will be presented in a separate paper (Sch\"oller et 
al., in preparation).

The mean longitudinal magnetic field, $\left< B_{\rm z}\right>$, is 
measured on the rectified and clipped spectra based on the relation
\begin{eqnarray} 
\frac{V}{I} = -\frac{g_{\rm eff}\, e \,\lambda^2}{4\pi\,m_{\rm e}\,c^2}\,
\frac{1}{I}\,\frac{{\rm d}I}{{\rm d}\lambda} \left<B_{\rm z}\right>\, ,
\label{eqn:vi}
\end{eqnarray} 

\noindent 
where $V$ is the Stokes parameter that measures the circular polarization, $I$
is the intensity in the unpolarized spectrum, $g_{\rm eff}$ is the effective
Land\'e factor, $e$ is the electron charge, $\lambda$ is the wavelength,
$m_{\rm e}$ is the electron mass, $c$ is the speed of light, 
${{\rm d}I/{\rm d}\lambda}$ is the wavelength derivative of Stokes~$I$, and 
$\left<B_{\rm z}\right>$ is the mean longitudinal (line-of-sight) magnetic field.

For the determination of the mean longitudinal stellar magnetic field, we
consider two sets of spectral lines: (i) the entire spectrum including all
available absorption and emission lines, apart from the usually strong
[\ion{O}{iii}] nebula emission lines near 5000\,\AA\ (in the following
referred to as wavelength set \emph{all}); and (ii) the subset of
spectral lines originating exclusively in the photosphere (and wind) of the
central star and not appearing in the surrounding nebula spectrum
(in the following wavelength set \emph{star}), thus
avoiding any potential contamination by nebular emission lines. This allows us
to get an idea of the impact of the presence of nebula emission lines and the
inclusion of the continuum windows on the magnetic field measurements. 
Details about the selected wavelength regions for the individual objects 
can be found in Appendix~\ref{A0}.

Note that we do not differentiate between 
absorption and emission lines, since the relation between the Stokes $V$ 
signal and the slope of the spectral line wing, as given by 
Eq.\,(\ref{eqn:vi}), holds for both type of lines, so that the signals 
of emission and absorption lines add up rather than cancel. For 
simplification, we assume a typical Land\'e  factor of 
$g_{\rm eff} \approx 1.2$ for all lines.

\begin{table*}
\caption{
Longitudinal magnetic field measurements of the central stars in our sample,
with the modified Julian date of mid-exposure given in Col.\,(2).
The mean longitudinal magnetic field measured with method R1 
(regression based on a single dataset) using the entire spectrum
(\emph{all}) and using only the uncontaminated stellar lines 
(\emph{star}), respectively, is presented in 
Cols.\,(3) and (5). The corresponding results based on method RM (regression 
based on $M=10^6$ statistical variations of the original dataset) are shown in 
Cols.\,(4) and (6). The average signal-to-noise ratio of the stacked Stokes\,$I$
spectrum is given in Col.\,(7). A magnetic field detection at a significance 
level of $3\,\sigma$ is indicated by the bold face entries in 
Cols.\,(3) -- (6). All quoted errors are $1\,\sigma$ uncertainties.
}
\label{tab:log_meas}
\centering
\begin{tabular}{lcr @{$\,\pm$} rr @{$\,\pm$} rr @{$\,\pm$} rr @{$\,\pm$} rcrc}
\hline
\hline\noalign{\smallskip}
\multicolumn{1}{l}{Name} &
\multicolumn{1}{c}{MJD} &
\multicolumn{4}{c}{$\left< B_{\rm z}\right>_{\rm all}$ [G]} &
\multicolumn{4}{c}{$\left< B_{\rm z}\right>_{\rm star}$ [G]} &
\multicolumn{1}{c}{} &
\multicolumn{1}{c}{S/N     } &
\multicolumn{1}{c}{Notes} \\
 &
 &
\multicolumn{2}{c}{R1} &
\multicolumn{2}{c}{RM} &
\multicolumn{2}{c}{R1} &
\multicolumn{2}{c}{RM} &
\multicolumn{1}{c}{} & 
 & \\
\noalign{\smallskip}\hline\noalign{\smallskip}
NGC\,246           & 55843.1712 &    $-$25  &      90  &  $-$21 & 79 &   $-$55 &  93 &   $-$51 &      78  &  & 1665 &  \\
                   & 55859.1612 &       62  &      77  &     62 & 75 &      66 &  79 &      66 &      76  &  & 1369 &  \\
\noalign{\smallskip}\noalign{\smallskip}
NGC\,1360          & 55839.2227 &   $-$ 7 &  97 &   $-$ 7 &  92 &      78 & 142 &           79 &     129  &  & 2161 &   \\ 
                   & 55843.3429 &  $-$109 & 145 &  $-$104 & 134 &  $-$192 & 207 &       $-$183 &     207  &  & 1952 &   \\
\noalign{\smallskip}\noalign{\smallskip}
NGC\,1514          & 55844.3521 &         138 &        94  &         139  &      88  &          122  &      94  &          123  &      89  &  & 2035 &   A-star \\
                   & 55859.2523 & $-${\bf 256} &  {\bf 85} & $-${\bf 257} & {\bf 77} &  $-${\bf 252} & {\bf 85} &  $-${\bf 252} & {\bf 77} &  & 2202 &  : \\
\noalign{\smallskip}\noalign{\smallskip}
IC\,418            & 55840.2816 & \multicolumn{8}{c}{only two sub-exposures,
  no measurement of $\left<B_{\rm z}\right>$} \\
                   & 55879.3257 & $-${\bf 181} & {\bf 54} &  $-${\bf 177} &  {\bf 54} & $-$204 &   97 &      $-$201  & 101   &  & 1870 &    \\
                   & 55899.1557 &      $-$157  &      64  &       $-$157  &       57  & $-$143 &   87 &      $-$143  &  83   &  & 1733 &    \\
\noalign{\smallskip}\noalign{\smallskip}
NGC\,2346          & 55885.2538 &      $-$44  &     74  &      $-$41  &     88  &       $-$31  &    104  &       $-$24  &    124  &  & 1264 & A-star \\
                   & 55905.3067 &      $-$86  &     45  &      $-$86  &     44  &       $-$54  &     59  &       $-$53  &     60  &  & 1860 &    :   \\
                   & 55906.2222 &      $-$82  &     62  &      $-$82  &     60  &       $-$23  &     68  &       $-$23  &     65  &  & 1723 &    :   \\
\noalign{\smallskip}\noalign{\smallskip}
NGC\,2392          & 55917.2243 &       46  &     73  &       46  &     78 &   $-$132  &    152  &  $-$135  &    145  &  & 1612 &    \\
                   & 55926.2776 &      137  &     81  &      136  &     87 &      219  &    163  &     224  &    133  &  & 1635 &    \\
\noalign{\smallskip}\noalign{\smallskip}
Hen\,2-36          & 55906.2864 &       $-$116 &      74 &       $-$116 &   69  &  $-$107 &  97 &    $-$107 &  93  &  & 1641 &  A-star \\ 
                   & 55908.3180 &        $-$99 &      92 &        $-$99 &   94  &   $-$70 & 111 &     $-$69 & 113  &  & 1434 &     :   \\
\noalign{\smallskip}\noalign{\smallskip}
LSS\,1362          & 55909.2720 &     38 &  236 &     35 &  234 &      157 & 263 &      152 & 264 &  &  985 & \\ 
                   & 55917.3188 &  $-$95 &  175 &  $-$91 &  177 &   $-$105 & 208 &   $-$101 & 208 &  & 1163 & \\
\noalign{\smallskip}\noalign{\smallskip}
NGC\,3132          & 55909.3216 &      38 &  70 &    38  &  70  &      97  &  86  &   97  &   87  &  & 1610 &  A-star \\  
                   & 55924.2987 &      65 &  51 &    65  &  60  &      74  &  62  &   73  &   76  &  & 2030 &     :   \\
\noalign{\smallskip}\noalign{\smallskip}
Hen\,2-108         & 55981.2533 &    261 &  126 &    253 &  135 &      374 &  179 &      351 &  213 &  & 881 & \\ 
                   & 55996.2235 & $-$118 &  142 & $-$115 &  121 &   $-$274 &  220 &   $-$280 &  218 &  & 882 & \\
                   & 56014.3206 &  $-$12 &  121 &   $-$1 &  140 &       98 &  201 &      115 &  175 &  & 804 & \\
\noalign{\smallskip}\noalign{\smallskip}
Hen\,2-113         & 55981.3145 &$-${\bf 58} & {\bf 18} &$-$58 & 24 & $-${\bf 78} & {\bf 25} &  $-${\bf 80} &  {\bf 26} &  & 1236 &  \\
                   & 55996.2851 &     $-$51  &      18  &$-$51 & 21 &      $-$47  &      24  &        $-$47 &       30  &  & 1316 &  \\
\noalign{\smallskip}\noalign{\smallskip}
Hen\,2-131         & 55996.3520 &  $-${\bf  92} & {\bf 29} &   $-$90 &  41 &   $-$109  &  41 &  $-$107 &  50 &  & 1972 &    \\ 
                   & 55997.3157 &  $-${\bf 120} & {\bf 32} &  $-$119 &  50 &    $-$90  &  39 &   $-$89 &  51 &  & 1936 &    \\
\noalign{\smallskip}\hline
\end{tabular}
\end{table*}

Given the Stokes $I$ and Stoke $V$ spectra of the selected wavelength region,
the mean longitudinal magnetic field $\left<B_{\rm z}\right>$ is derived by 
linear regression employing two different methods, 
in the following referred to as method R1 and RM. 

In \emph{method R1}, $\left<B_{\rm z}\right>$ is defined by the slope of the 
weighted linear regression line through the measured data points, where
the weight of each data point is given by the squared signal-to-noise ratio
of the Stokes $V$ spectrum. The formal $1\,\sigma$ error of 
$\left<B_{\rm z}\right>$ is obtained from the standard relations for weighted 
linear regression. This error is inversely proportional to the rms  
signal-to-noise ratio of Stokes $V$. Finally, the factor
$\sqrt{\chi^2_{\rm min}/\nu}$ is applied to the error determined from the 
linear regression, if larger than 1 (for details see Appendix~\ref{A11}).

In \emph{method RM}, we generate $M=10^6$ statistical variations of the 
original dataset by the bootstrapping technique, and analyze the resulting 
distribution $P(\left<B_{\rm z}\right>)$ of the $M$ regression results. Mean 
and standard deviation of this distribution are identified with the most 
likely mean longitudinal magnetic field and its $1\,\sigma$ error, 
respectively. The main advantage of this method is that it provides an 
independent error estimate (see Appendix~\ref{A1M}). 

Normalized FORS\,2 Stokes~$I$ spectra of all sample stars 
(averaged over the different epochs)
are displayed in Fig.~\ref{fig:norm}; line identifications of
well-known spectral lines are listed at the top. 
The appearance of the spectra already provides a clue of the binarity status
of the central stars: 
the spectra of NGC\,1514, NGC\,2346, Hen\,2-36, and NGC\,3132
are largely dominated by companions of spectral type A. The spectra of the
central stars Hen\,2-131 and Hen\,2-113, and to a lesser extent Hen\,2-108,
predominantly show emission lines, whereas the remaining central stars show a
predominance of absorption lines.
 
The results of our stellar magnetic field measurements are presented in 
Table~\ref{tab:log_meas}. For each object 
and each observing night, the mean longitudinal magnetic field,
as derived from methods R1 and RM, is listed for both sets of lines 
(\emph{all} and \emph{star}). The mean longitudinal 
magnetic field derived from the original dataset with method R1 
is always close to the result obtained with the Monte-Carlo method RM, 
$\overline{\left<B_{\rm z}\right>}$. Moreover, the $1\,\sigma$ uncertainties 
obtained from method R1 (Eq.\,\ref{sigmaR1}) and method 
RM (Eq.\,\ref{sigmaRM}), respectively, agree remarkably well.
No significant fields were detected in the null spectra calculated using the
formalism described by Bagnulo et al.\ (\cite{Bagnulo2012}).

Considering the entire spectrum (wavelength set \emph{all}),
we achieve formal $3\,\sigma$ detections for NGC\,1514 (second night) 
and IC\,418 (second night) with both methods (R1 and RM), and for 
the WR-type central star Hen\,2-113 and the `weak emission line star' 
(WELS) Hen\,2-131 with method R1 only.
Repeating the magnetic field measurements on the restricted
wavelength range with clean stellar lines only (wavelength set \emph{star}),
formal $3\,\sigma$ detections are confirmed in the case of NGC\,1514 (second
night) and Hen\,2-113 (first night). For IC\,418 and Hen\,2-131, the
$\left<B_{\rm z}\right>$ values found from wavelength sets \emph{all} and
\emph{star} are very similar, but the formal errors are larger in the latter
case, resulting in $2\,\sigma$ detections only.

NGC\,1514 is the only object where we detect a magnetic field 
consistently with both methods and with both sets of lines above the
$3\,\sigma$ significance level, indicating a mean longitudinal field 
of $\approx 250$~G on the second night.
Note, however, that in NGC\,1514 the spectrum of the central star is 
dominated by an A-type companion, such that we only measure the magnetic 
field of the A-type star, while the magnetic field of the hot, compact 
central star proper remains inaccessible.

The magnetic field detection in the WR-type central star 
Hen\,2-113 is significant above the $3\,\sigma$ level with both R1 and
RM when considering clean stellar lines only. However, the detected magnetic
field is very weak, $\left<B_{\rm z}\right> \approx 80$~G.


In the following section, we discuss the individual targets of our sample
in more detail.

\section{Individual targets: general information, spectral variability, 
magnetic field detection, X-rays}
\label{sect:individual}

For all our targets, the basic data of the central stars and their nebulae, as
collected from the literature, are summarized in Tables\,\ref{tab:objects} and
\ref{tab:cspn_dat}, together with related references. For most of
the targets, additional information may be found in Table~1 of Kastner et 
al.\ (\cite{Kastner2012}).

\paragraph{\it NGC\,246:}
The hot central star ($T_{\rm eff}\approx 150$~kK) belongs to the group of 
PG\,1159 stars, which are hot hydrogen deficient post-AGB stars 
(e.g., Werner \& Herwig \cite{WernerHerwig2006}). In the Hertzsprung-Russell 
diagram, they cover a region comprising the hottest central stars of planetary 
nebulae and white dwarfs ($T_{\rm eff} = 75$ to $200$\,kK, 
$\log g=5.5$ to $8.0$). 
Their H deficiency is most probably the result of a late He-shell flash. Their 
envelopes are mainly composed of He, C, and O, with rather diverse abundance 
patterns (mass fractions: He = $0.30$ to $0.85$, C = $0.13$ to $0.60$, 
O = $0.02$ to $0.20$).

The projected rotational velocity of the central star was determined
spectroscopically by Rauch et al.\ (\cite{Rauch2003}) as 
$v\,\sin i = 77^{+23}_{-17}$~km/s.
NGC\,246 is a round planetary nebula whose central star is a known visual 
binary, with a companion of spectral type G8-K0~V 
(e.g., De~Marco \cite{DeMarco2009}). 
Chandra observations revealed a central point source of 
soft X-rays (for details see Hoogerwerf et al.\ \cite{Hoogerwerf2007}).

We observed this central star on two different nights, without 
detecting any spectral variability in the FORS\,2 spectra. 
No magnetic field was detected on either night. The error estimates
of both methods (R1 and RM) are in good agreement, yielding 
$\sigma_B \la 100$~G, irrespective of the wavelength set 
considered for the magnetic field measurement. 
Hence, the $3\,\sigma$ upper limit for the mean longitudinal magnetic 
field is roughly $|\left<B_{\rm z}\right>| \la 300$~G.

\paragraph{\it NGC\,1360:}
The central star of this elliptical PN has recently been analyzed by
Herald \& Bianchi (\cite{Herald2011}). According to this study, the
mass loss rate is the lowest of all central stars of our sample (see
Tab.\,\ref{tab:cspn_dat}). Kastner et al.\ (\cite{Kastner2012}) find
this central star to be a point source of soft X-rays.

Based on spectropolarimetric data obtained with FORS\,1, Jordan 
et al.\ (\cite{Jordan2005}) reported the detection of a magnetic field of the
order of several kG.  Leone et al.\ (\cite{Leone2011,Leone2014}) 
re-observed the central
star with FORS\,2, but could not confirm the presence of a magnetic field
within an uncertainty of $\sim$100\,G.  Later on, Jordan et
al.\ (\cite{Jordan2012}) reanalyzed their own data and concluded that no
magnetic field is present in this star.  Our observations with FORS\,2 on two
different nights with an uncertainty $\approx 150$\,G are fully in agreement
with these recent works. The $3\,\sigma$ upper limit for the mean 
longitudinal magnetic field is $|\left<B_{\rm z}\right>| \la 450$~G.


\begin{figure}
\centering
\mbox{\includegraphics[bb=28 28 580 350,width=0.48\textwidth]
{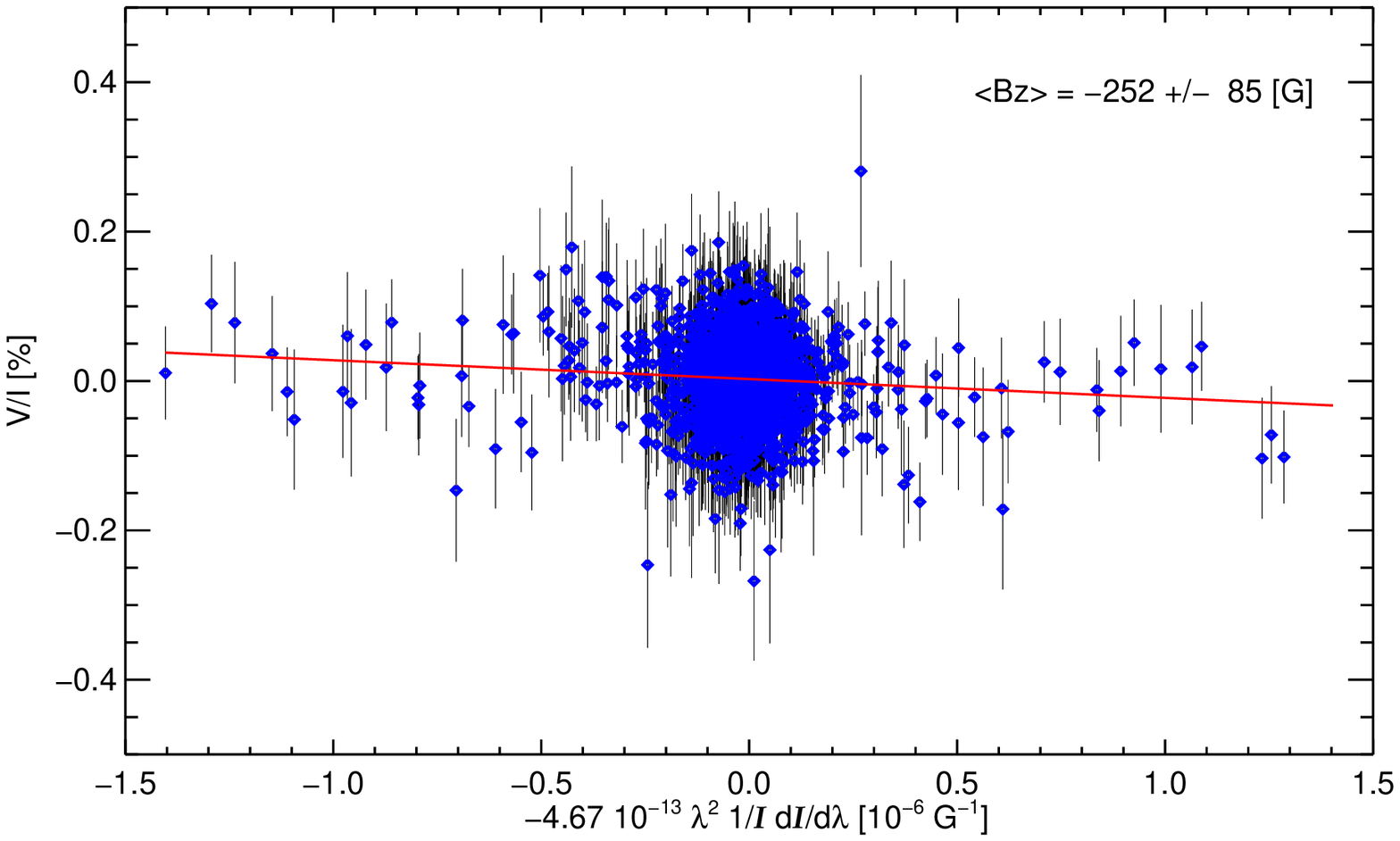}}
\mbox{\includegraphics[bb=0 0 360 216,width=0.48\textwidth]
{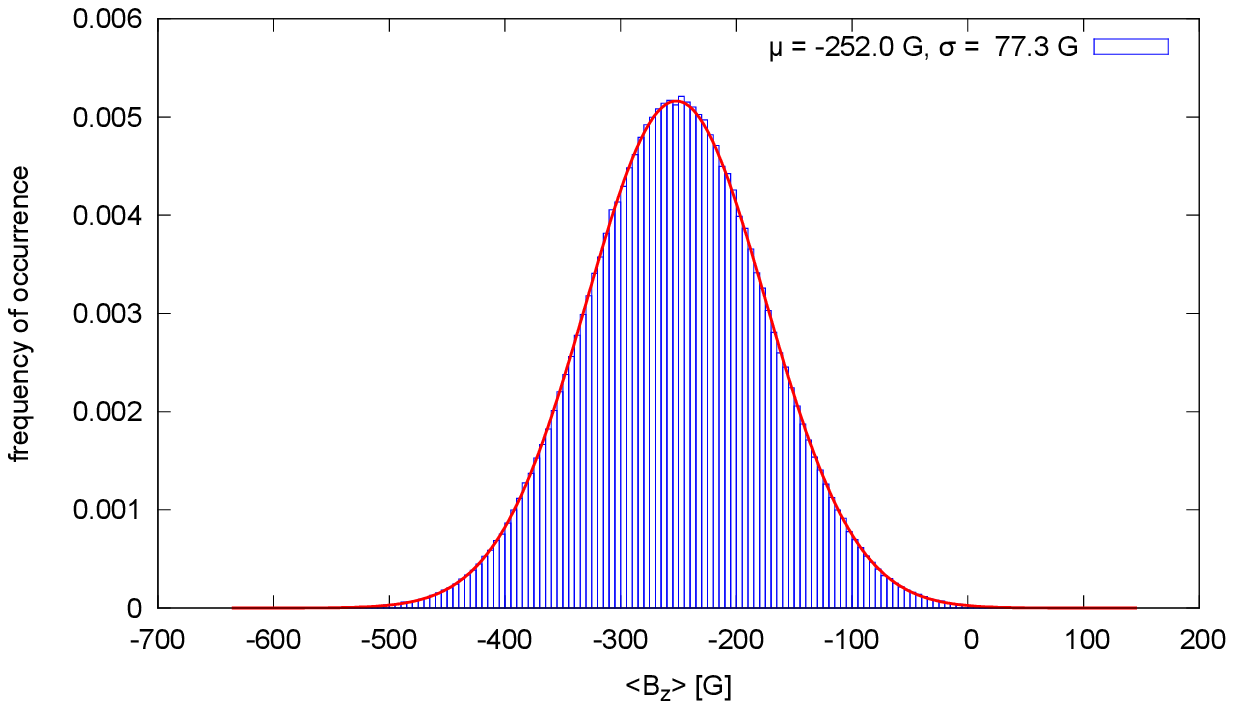}}
\caption{
NGC\,1514: Regression detection of a magnetic field
on the second night, using uncontaminated stellar lines only. 
Methods R1 and RM yield a mean longitudinal magnetic field of 
$\left<B_{\rm z}\right>=-252\pm85$\,G (top) and 
$\left<B_{\rm z}\right>=-252\pm77$\,G (bottom), respectively.
In the bottom panel, and the following figures of this kind, the histogram 
plot has a bin size of $5$\,G, and the (red) curve represents a Gaussian with 
the same mean value and standard deviation as the histogram data.}
\label{fig:ngc1514}
\end{figure}

\paragraph{\it NGC\,1514:}
The spectrum of the central star is largely dominated by a
companion of spectral type A0~III (e.g., De~Marco \cite{DeMarco2009}).
Recent Chandra observations revealed that this elliptical PN harbors a 
central point source of hard X-rays (Kastner et al.\ \cite{Kastner2012}).

The inspection of our Stokes~$I$ spectra obtained on two different nights
indicates a very low, if not negligible, spectral variability.


On the first night, no significant magnetic field is detectable
(${\left<B_{\rm z}\right> \approx 130\pm90}$\,G). For the second epoch,
however, the magnetic field measurements showed a detection 
of a longitudinal field of negative polarity at a $3\,\sigma$
significance level (${\left<B_{\rm z}\right>=-250\pm80}$\,G), using
the clean stellar lines only. The results obtained from methods R1 
and RM are fully consistent, as illustrated in Fig.\,\ref{fig:ngc1514}:
the top panel shows the linear regression plot of method R1, $V/I$ versus
$-(g_{\rm eff}\, e\,\lambda^2)/(4\pi\,m_{\rm  e}\,c^2)\, 
I^{-1}\,{\rm d}I/{\rm d}\lambda$, where the slope represents 
$\left< B_{\rm z}\right>$, while the bottom panel displays the distribution 
of $\left< B_{\rm z}\right>$ as obtained with method RM. We find similar
results for the magnetic field if we perform our analysis on the entire 
spectrum, which is not surprising, however, since there are only a few 
nebular emission lines that contaminate the stellar spectrum.

Based on this clear detection, the lower limit of the magnetic field 
strength is  $|\left<B_{\rm z}\right>| \ga 250$~G. However, the field must be 
assigned to the A-type companion of the central star proper.

\begin{figure*}
\centering
\mbox{\includegraphics[bb=14 28 580 354,clip=true,width=0.48\textwidth]
{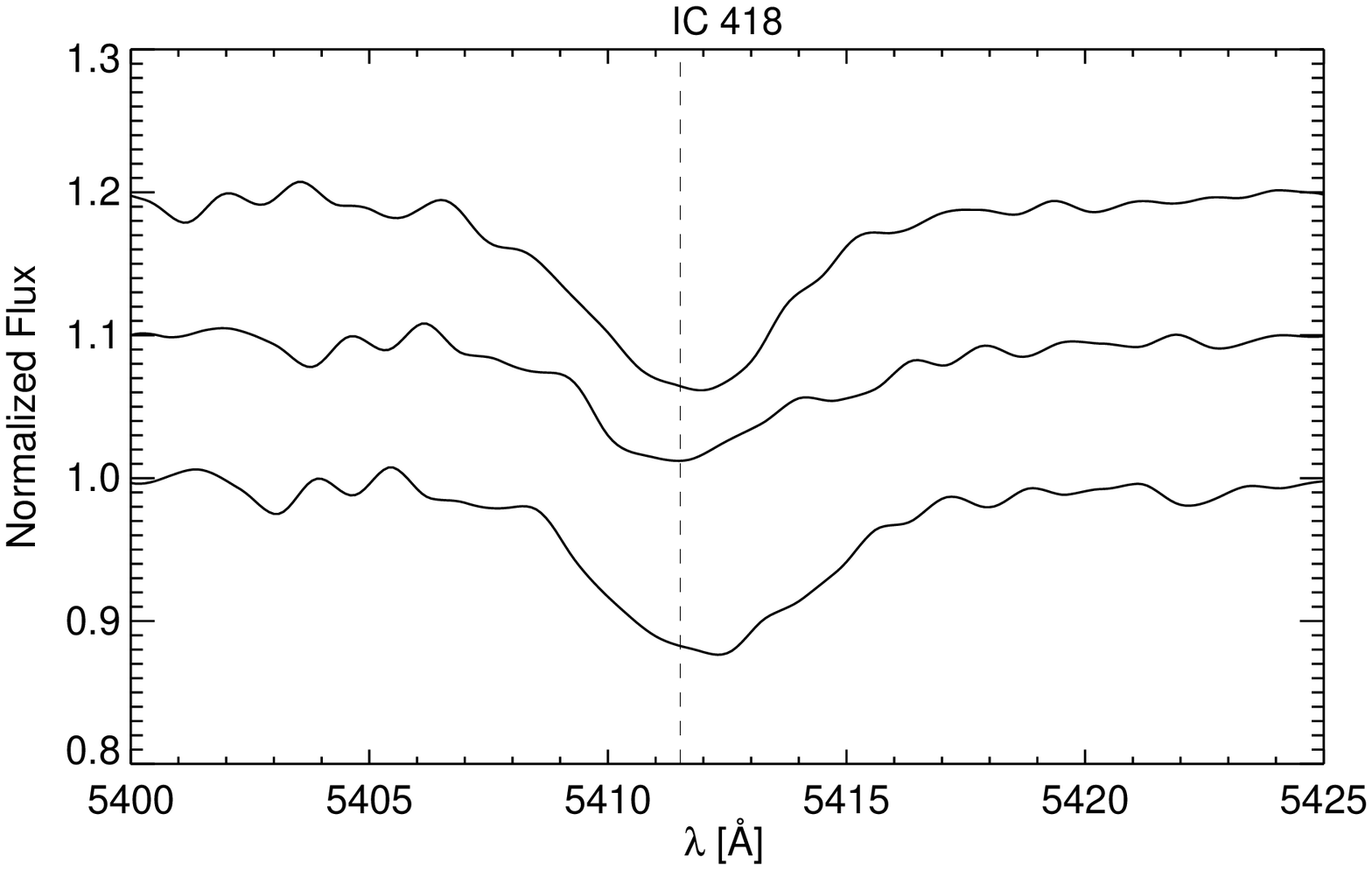}}
\mbox{\includegraphics[bb=14 28 580 354,clip=true,width=0.48\textwidth]
{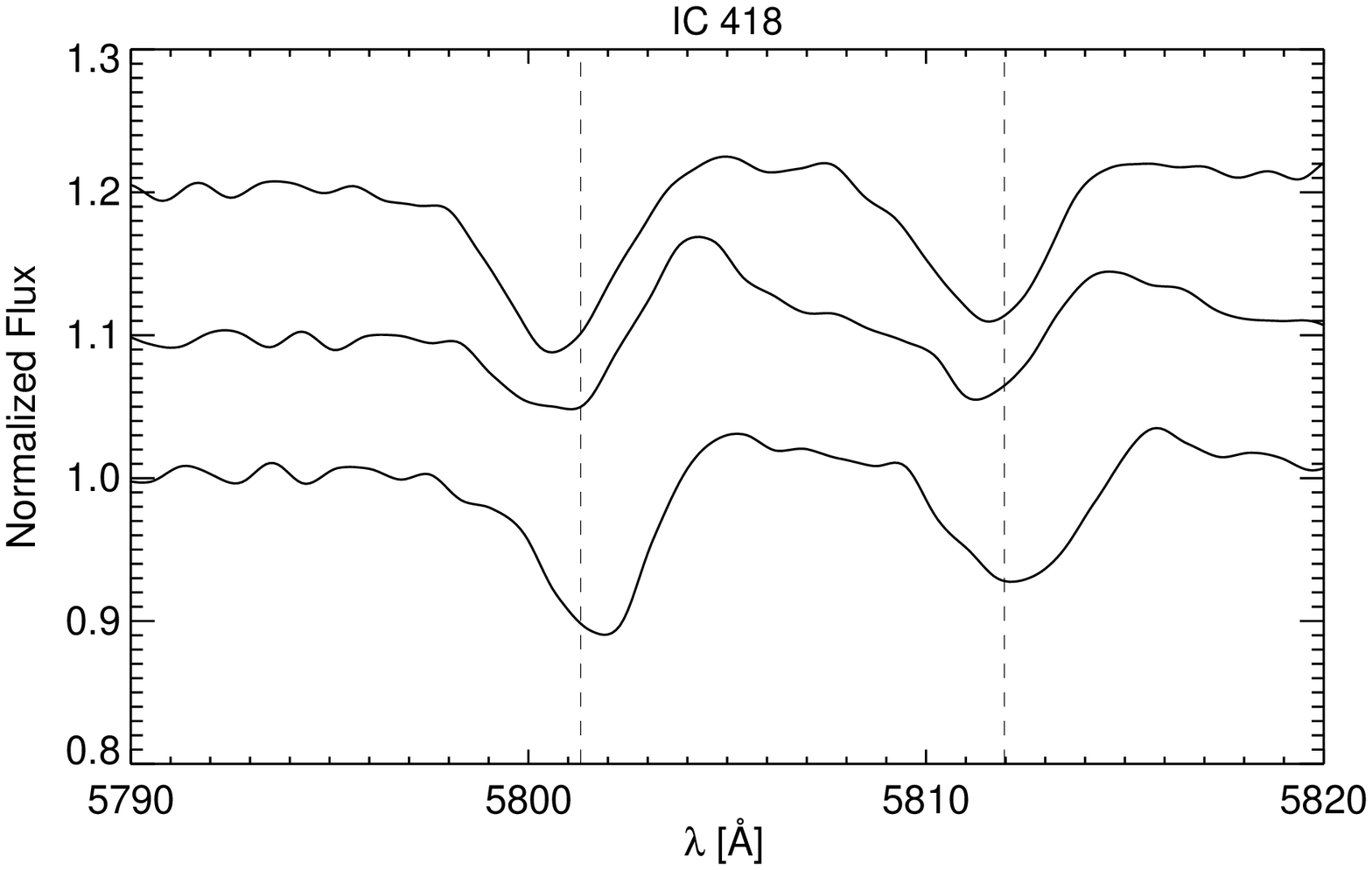}}
\caption{
IC\,418: Normalized Stokes~$I$ spectra of the central star in the spectral
regions around the \ion{He}{ii} line at $\lambda$\, 5411.5\,\AA\ (left) 
and the \ion{C}{iv} lines $\lambda$\,5801.3 and 5812.0\,\AA\ (right) obtained 
on three different nights. The spectra are shifted in vertical direction 
by 0.1 units for clarity, with the epoch increasing from bottom to top 
(the ordinate is correct for the lowermost spectrum).}
\label{fig:ic418_i}
\end{figure*}

\begin{figure}
\centering
\mbox{\includegraphics[bb=28 28 580 350,width=0.48\textwidth]
{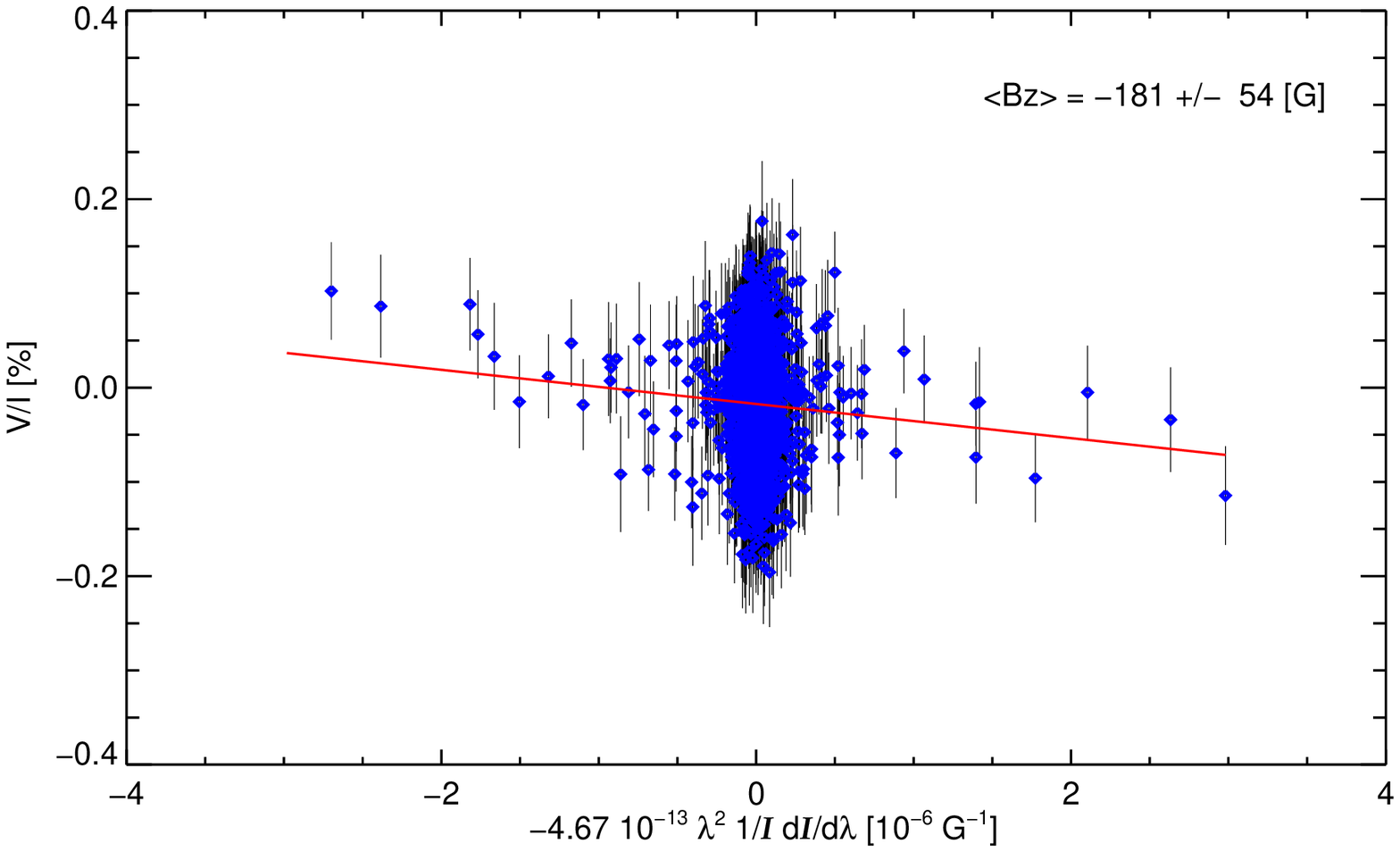}}
\mbox{\includegraphics[bb= 0  0 360 216,width=0.48\textwidth]
{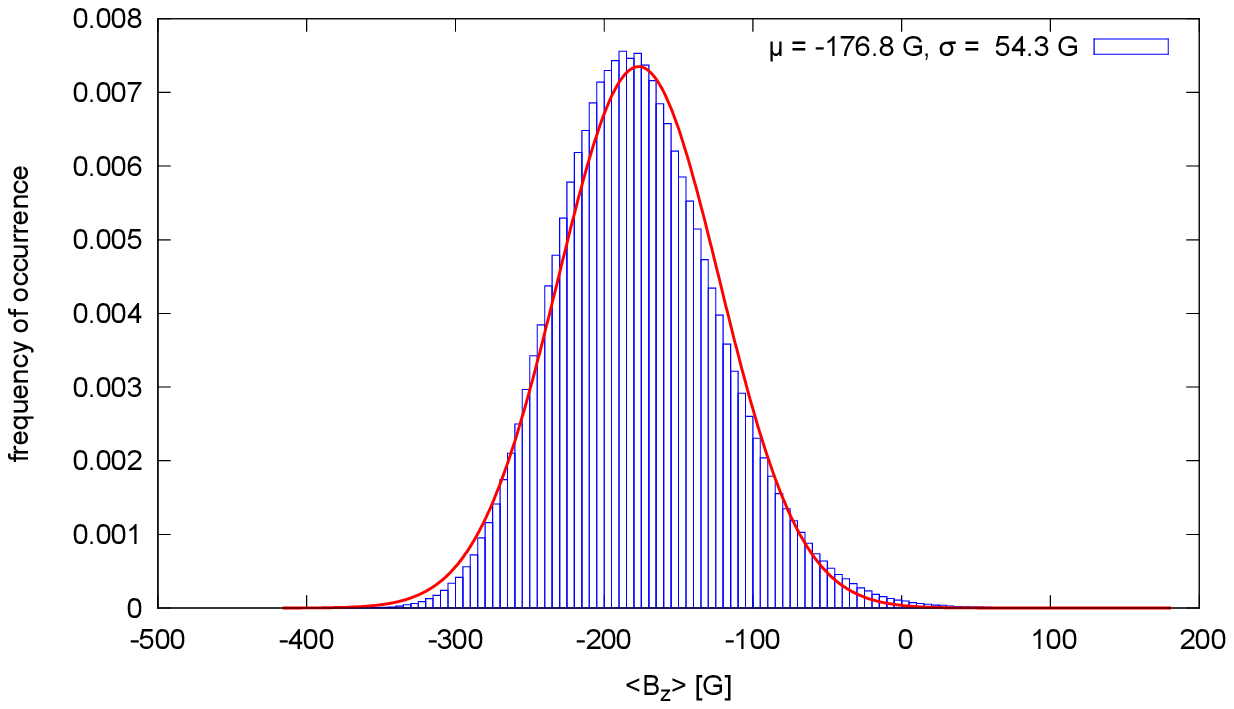}}
\caption{
IC\,418: Regression detection with method R1 of a 
$\left<B_{\rm z}\right>=-181\pm54$\,G mean longitudinal magnetic field
on the second night, using all lines (top). The corresponding 
$\left<B_{\rm z}\right>$ distribution obtained from method RM (bottom) 
deviates from a Gaussian (red curve) and indicates 
$\left<B_{\rm z}\right>=-177\pm54$\,G.}
\label{fig:ic418_b}
\end{figure}

\paragraph{\it IC\,418:}
The central star of this young, elliptical planetary nebula was 
analyzed extensively (see, e.g.,
Mendez at al.\ \cite{Mendez1988}, \cite{Mendez1992}; Kudritzki et al.\
\cite{Kudritzki1997,Kudritzki2006}; Pauldrach et al.\ \cite{Pauldrach2004}; 
Morisset \& Georgiev \cite{Morisset2009}). The nebula 
is listed as a source of diffuse X-ray emission in Table~1 of Kastner 
et al.\ (\cite{Kastner2012}). A more detailed recent analysis of the 
X-ray properties of IC\,418 is listed in Ruiz et al.\ (\cite{Ruiz2013}). 
The projected rotational velocity of the central star 
was recently determined from a Fourier transform analysis of photospheric
lines by Prinja et al.\ (\cite{Prinja2012}), who found $v\,\sin i = 56$~km/s.

Spectropolarimetric observations of the central star were carried out on three
different nights. A significant spectral variability is detected in 
the \ion{He}{ii} lines at $\lambda$\,4541 and 5412\,\AA\ and in the \ion{C}{iv}
lines near $\lambda$\,5800\,\AA. Note that these lines appear not to be 
blended with nebular emission lines (see also Fig.\,1 of Mendez et al.\ 
\cite{Mendez1988}).
In Fig.~\ref{fig:ic418_i}, we present as an example 
the variability of the \ion{He}{ii} line at $\lambda$\,5412\,\AA\ and 
of the \ion{C}{iv} line profiles near $\lambda$\,5800\,\AA.

A mean longitudinal magnetic field of negative polarity was detected at the
$3\,\sigma$ level of significance on the second night when using the entire 
spectrum: ${\left<B_{\rm z}\right>=-181\pm54}$\,G (method R1) and
${\left<B_{\rm z}\right>=-177\pm54}$\,G (method RM), as shown in 
Fig.\,\ref{fig:ic418_b}. The results obtained from both methods
are essentially identical. However, when using the uncontaminated
stellar lines only, the formal error increases to 
$\sigma_{\rm B} \approx 100$~G, and we are left with a $2\,\sigma$ signal
in this case. It is noteworthy that all measurements for this object fall
in the narrow range $175 \pm 30$\,G. Nevertheless, adopting $100$~G as 
the valid error estimate, the magnetic field detection in IC\,418 must 
be considered as marginal and we infer a conservative $3\,\sigma$ upper 
limit for the mean longitudinal magnetic field of 
$|\left<B_{\rm z}\right>| \la 300$~G.

\paragraph{\it NGC\,2346:}
The optical spectrum of the central star of this bipolar PN is largely 
dominated by a companion of spectral type A5~V (e.g., De~Marco 
\cite{DeMarco2009}).
The inspection of the Stokes~$I$ spectra obtained on three different nights
indicates a very low spectral variability, as in the case of NGC\,1514.
Higher resolution spectra would be needed to establish a possible
low-amplitude variability of the A-type stellar spectrum.

None of our spectropolarimetric measurements indicates the presence
of a magnetic field signal.
We conclude that the mean longitudinal magnetic field of the A-type 
companion, $|\left<B_{\rm z}\right>|$, does not exceed $300$~G.

According to Kastner et  al.\ (\cite{Kastner2012}), neither the central
star (wind) nor the central cavity of the PN shows detectable X-ray emission.

\paragraph{\it NGC\,2392:}
The spectrum of the central star of this round/elliptical PN was 
analyzed extensively 
(see, e.g., Mendez at al.\ \cite{Mendez1988}, \cite{Mendez1992}; 
Kudritzki et al.\ \cite{Kudritzki1997,Kudritzki2006}; 
Pauldrach et al.\ \cite{Pauldrach2004}). According to Kastner et 
al.\ (\cite{Kastner2012}), the central cavity of the nebula is
a source of diffuse X-ray emission and, in addition, harbors a hard X-ray point
source. A more detailed recent analysis of the X-ray properties of NGC\,2392 
is presented by Ruiz et al.\ (\cite{Ruiz2013}). Possibly, the central star has
a late-type (dM) companion (Ciardullo et al.\ \cite{Ciardullo1999}).

No significant line profile variability is detected, neither in 
the Stokes~$I$ nor in the Stokes~$V$ spectra. Similarly, 
no significant magnetic field signal was detected on either observing
night. Typical $1\,\sigma$ errors are $\approx 80$~G when the entire spectrum 
is used, and $\approx 150$~G when only the clean stellar lines are
considered for the magnetic field measurement. A conservative
$3\,\sigma$ upper limit for the mean longitudinal magnetic field of the 
central star of NGC\,2392 is therefore $|\left<B_{\rm z}\right>| \la 450$~G.

\begin{figure*}[hbt]
\centering
\mbox{\includegraphics[bb=14 28 580 354,clip=true,width=0.48\textwidth]
{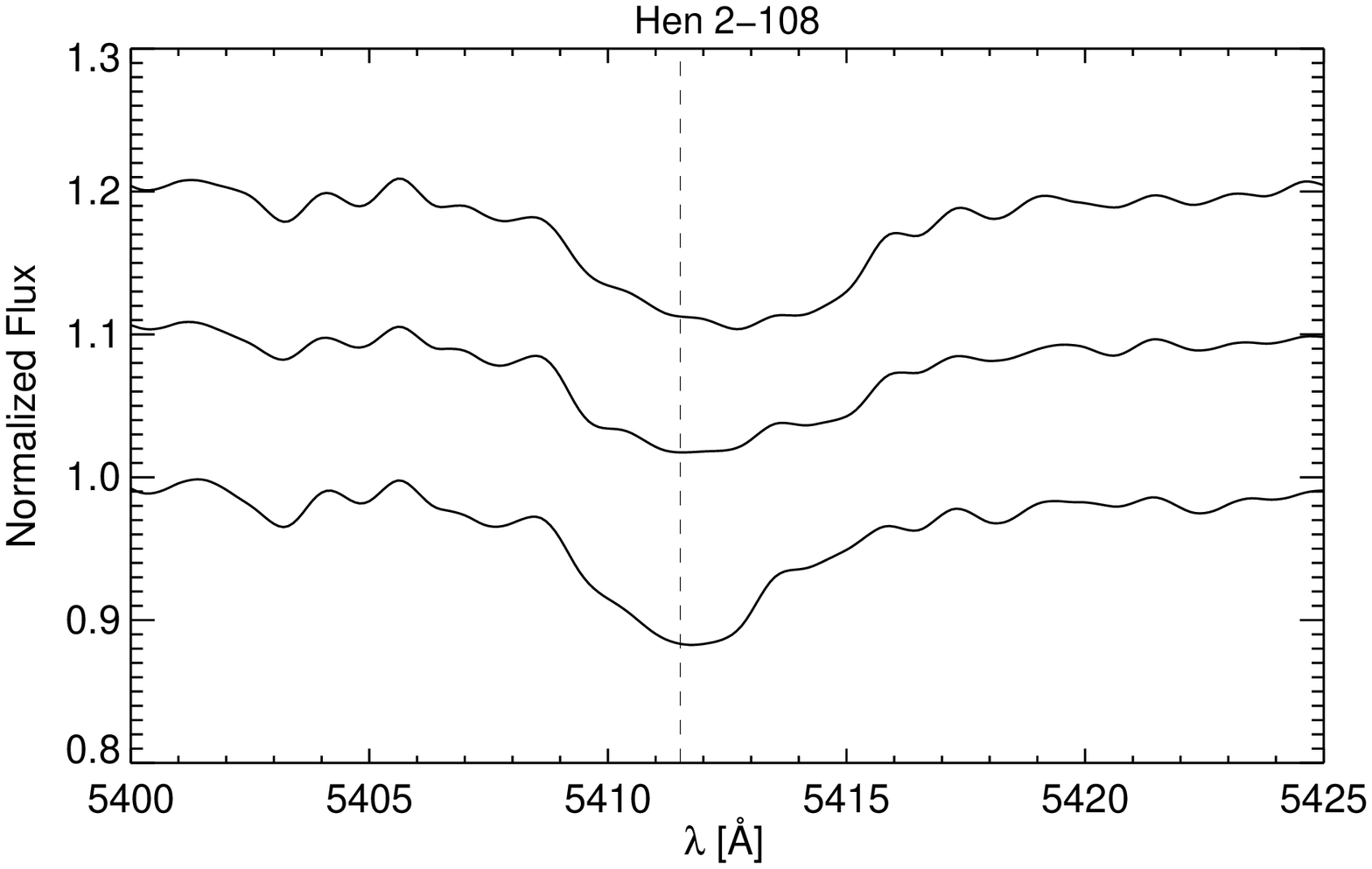}}
\mbox{\includegraphics[bb=14 28 580 354,clip=true,width=0.48\textwidth]
{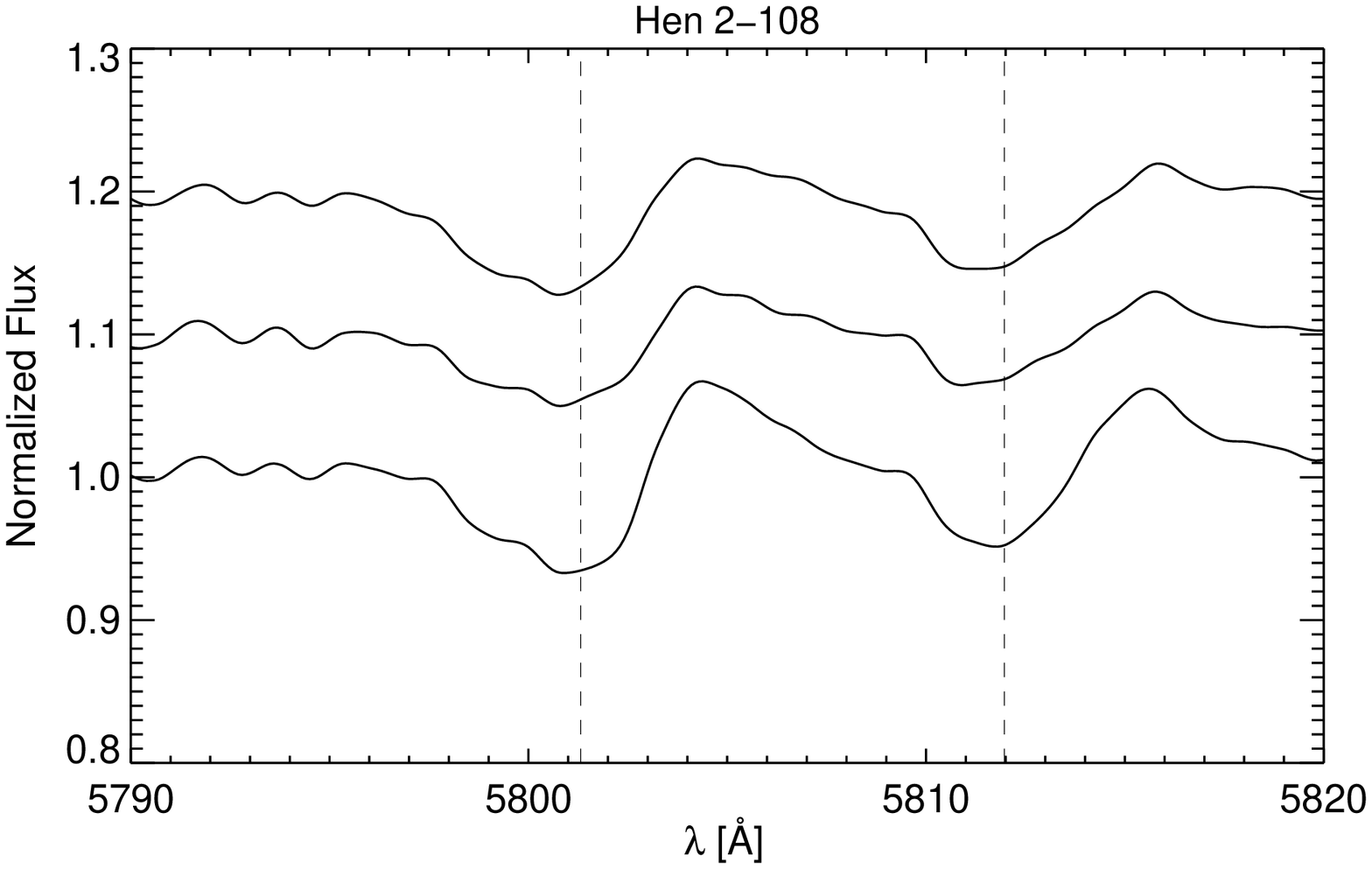}}
\caption{
Hen\,2-108: Normalized Stokes~$I$ spectra of the central star of the round
planetary nebula in the spectral
regions around the \ion{He}{ii} line at $\lambda$\, 5411.5\,\AA\ (left) 
and the \ion{C}{iv} doublet at $\lambda$\,5801.3 and 5812.0\,\AA\ (right) 
obtained on three different nights. The spectra are shifted in vertical 
direction by 0.1 units for clarity, with the epoch increasing from bottom 
to top.}
\label{fig:hen2_108}
\end{figure*}

\paragraph{\it Hen\,2-36:}
The optical spectrum of the central star of this bipolar PN is largely 
dominated by a companion of spectral type A. Spectral variability is very 
weak in both Stokes~$I$ and Stokes~$V$ spectra.
Our measurements on two different observing nights show no
evidence for the presence of a magnetic field in the central star of 
Hen\,2-36, neither with method R1 nor with method RM, and irrespective
of the considered spectral range. With typical errors of $\sigma_{\rm B}
\approx 100$~G, we estimate a $3\,\sigma$ upper limit of 
$|\left<B_{\rm z}\right>| \la 300$~G.

\paragraph{\it LSS\,1362:}
Using spectropolarimetric data obtained with FORS\,1, Jordan et
al.\ (\cite{Jordan2005}) reported the detection of a magnetic field of the
order of several kG.  Leone et al.\ (\cite{Leone2011,Leone2014}) reobserved 
the central
star with FORS\,2 but, similar to the case of NGC\,1360, the authors could
not confirm the presence of a magnetic field within an uncertainty of
$\sim$290\,G. Likewise, Jordan et al.\ (\cite{Jordan2012}) reanalyzed their
own data and concluded that no magnetic field is detectable in this star. 
As with NGC\,1360, our observations with FORS\,2 on two different nights,
with an uncertainty of about 300\,G, do not indicate the presence of a
magnetic field, in full agreement with these two recent works. 
A conservative $3\,\sigma$ upper limit for the mean longitudinal field
of this central star is estimated to be $|\left<B_{\rm z}\right>| \la 800$~G.


\paragraph{\it NGC\,3132:}
The optical spectrum of the central star of this elliptical PN is largely 
dominated by a companion of spectral type A2~IV-V (e.g., De~Marco 
\cite{DeMarco2009}). The inspection of the Stokes~$I$ spectra obtained on 
two different nights indicates a very low spectral variability, as in the 
other targets with A-type companions. 

No significant ($\ga 3\,\sigma$ for both sets of lines) mean longitudinal 
magnetic field  was measured on either night, even though the error estimates 
of both methods (R1 and RM) are less than $100$~G. The estimated $3\,\sigma$ 
upper limit for the mean longitudinal magnetic field of the A star companion 
is therefore relatively low, $|\left<B_{\rm z}\right>| \la 270$~G.

No X-ray detection is reported for NGC\,3132 by Kastner et al.\ 
(\cite{Kastner2012}), neither from the planetary nebula nor from its 
central star.

\paragraph{\it Hen\,2-108:}
The spectrum of the central star of this round PN appears very similar to 
that of IC\,418 (cf.\ Fig.\,\ref{fig:norm}, and Fig.\,1 of 
Mendez et al.\ \cite{Mendez1988}). 
According to Tylenda et al.\ (\cite{Tylenda1993}), the central star of 
Hen\,2-108 is a weak emission line star (WELS). Such stars have normal 
chemical composition, but show narrow lines from the $\lambda$\,4650
\ion{N}{iii}--\ion{C}{iii}--\ion{C}{iv} blend and the \ion{C}{iv}
$\lambda$\,5806 doublet;  \ion{C}{iii} $\lambda$\,5696 is absent or weak, while
\ion{He}{ii} $\lambda$\,4686 emission is often seen. These lines are also seen
in some massive O-type stars, low-mass X-ray binaries, and cataclysmic
variables. Further analyses of the central star can be found in Mendez et al.\ 
(\cite{Mendez1988,Mendez1992}) and Pauldrach et al.\ (\cite{Pauldrach2004}).
No information is available concerning the X-ray emission of this PN.

The Stokes~$I$ spectra show some variability, as demonstrated in 
Fig.~\ref{fig:hen2_108}, where we display the variable \emph{stellar} line 
profiles of \ion{He}{ii} $\lambda$\, 5412\,\AA\ and the \ion{C}{iv} lines 
near $\lambda$\,5800\,\AA.
However, we do not detect any mean longitudinal magnetic field at a 
significance level of $3\,\sigma$ on any of the three nights, irrespective 
of the adopted method and set of lines. Accepting the larger error 
estimates derived with the restricted wavelength set \emph{star},
we obtain a conservative $3\,\sigma$ upper limit for the mean longitudinal 
field of this central star of $|\left<B_{\rm z}\right>| \la 700$~G.

\begin{figure}
\centering
\mbox{\includegraphics[bb=28 28 580 350,width=0.48\textwidth]
{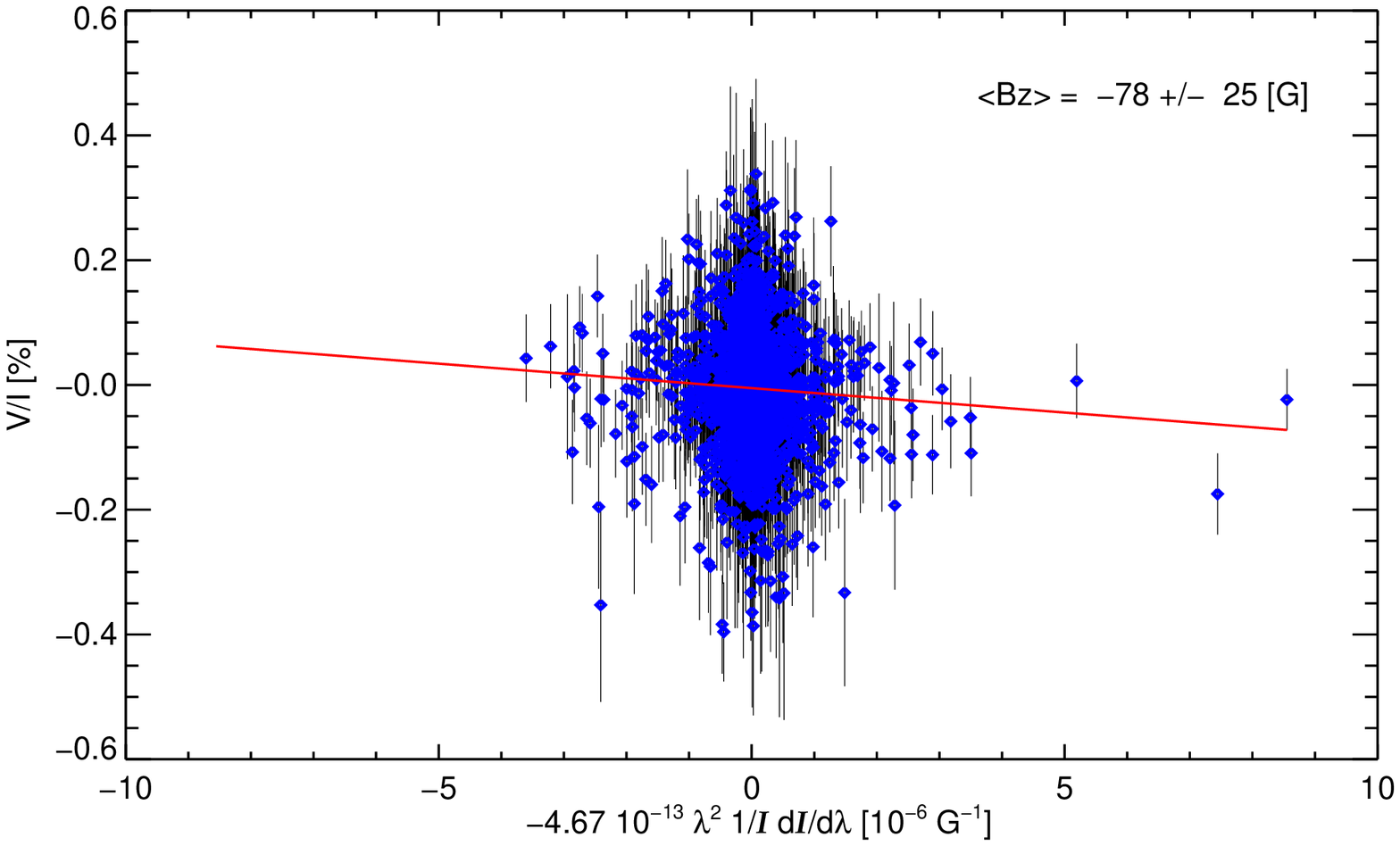}}
\mbox{\includegraphics[bb= 0  0 360 216,width=0.48\textwidth]
{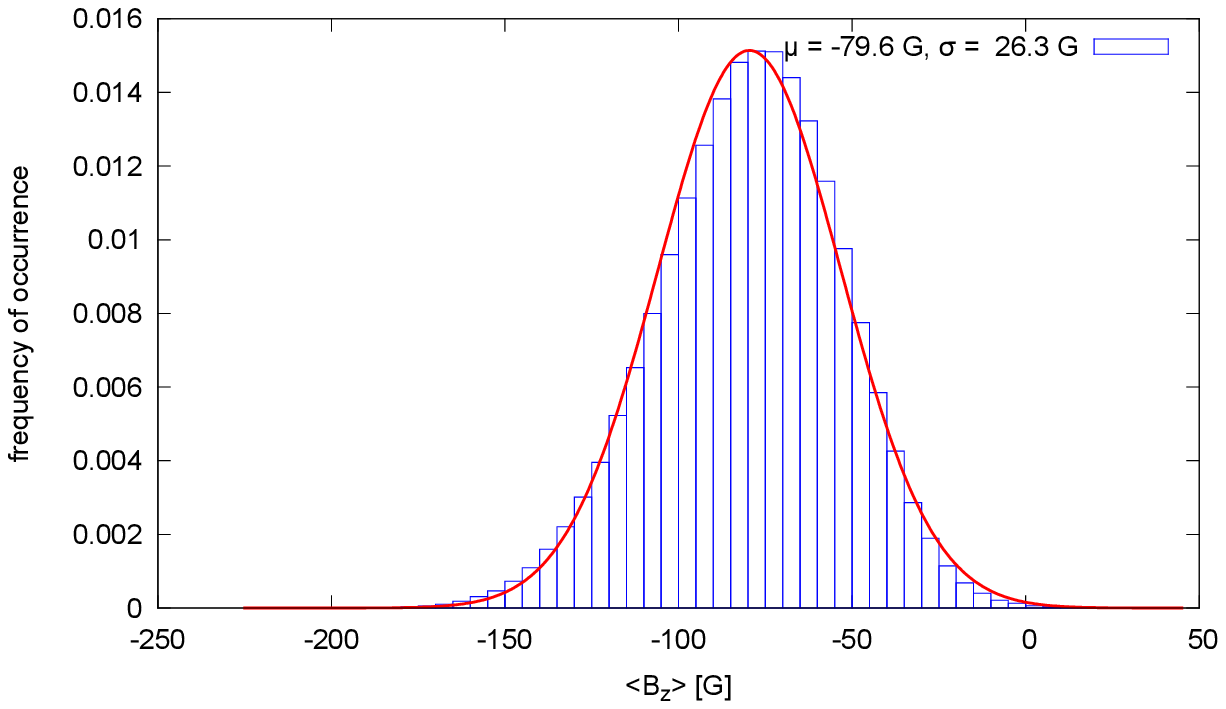}}
\caption{
Hen\,2-113: Regression detection of a $\left<B_{\rm z}\right>=-78\pm25$\,G 
mean longitudinal magnetic field on the first night with method R1, 
using uncontaminated stellar lines only (top), and corresponding 
(slightly non-Gaussian) distribution of $\left<B_{\rm z}\right>$ obtained 
from method RM (bottom).}
\label{fig:hen2_113_b}
\end{figure}

\begin{figure}
\centering
\mbox{\includegraphics[bb=14 28 580 354,clip=true,width=0.48\textwidth]
{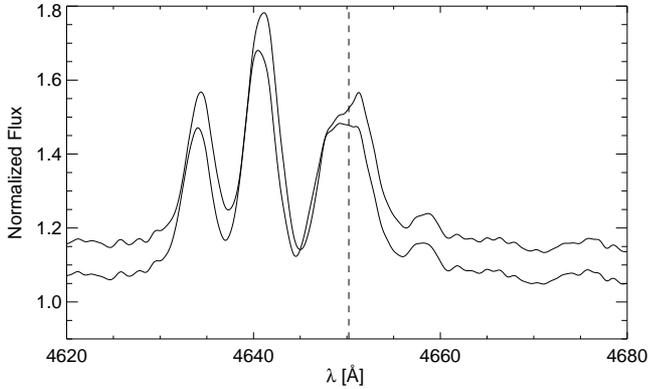}}
\caption{
Hen\,2-131: Normalized Stokes~$I$ spectra of the central star in the
spectral region around the \ion{C}{iii} feature at
$\lambda$\,4650.2\,\AA, obtained on two different nights.
The spectrum of the second night is shifted upward in vertical direction 
by 0.1 units for clarity.}
\label{fig:hen2_131_i}
\end{figure}

\begin{figure}
\centering
\mbox{\includegraphics[bb=28 28 580 350,width=0.48\textwidth]
{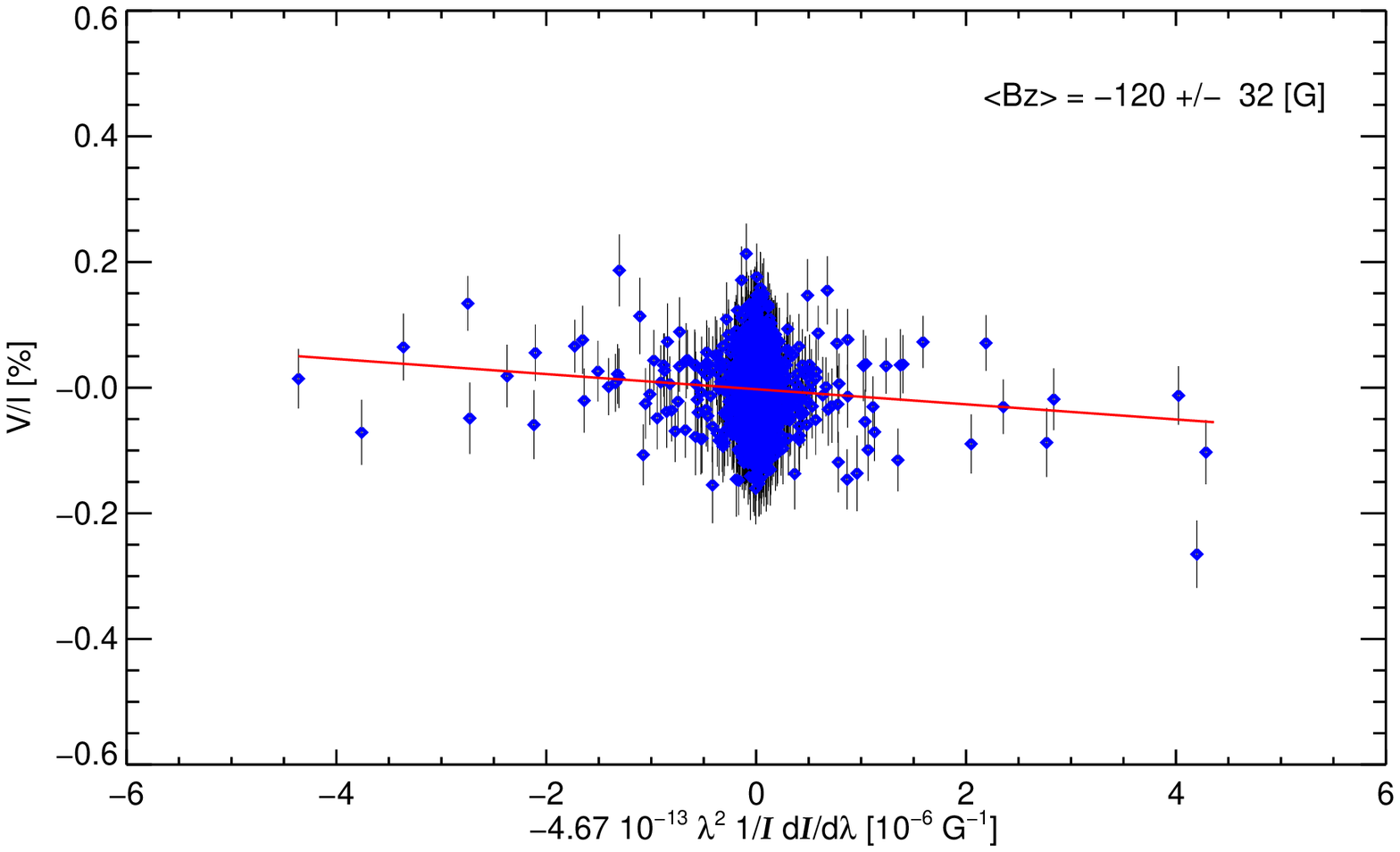}}
\mbox{\includegraphics[bb= 0  0 360 216,width=0.48\textwidth]
{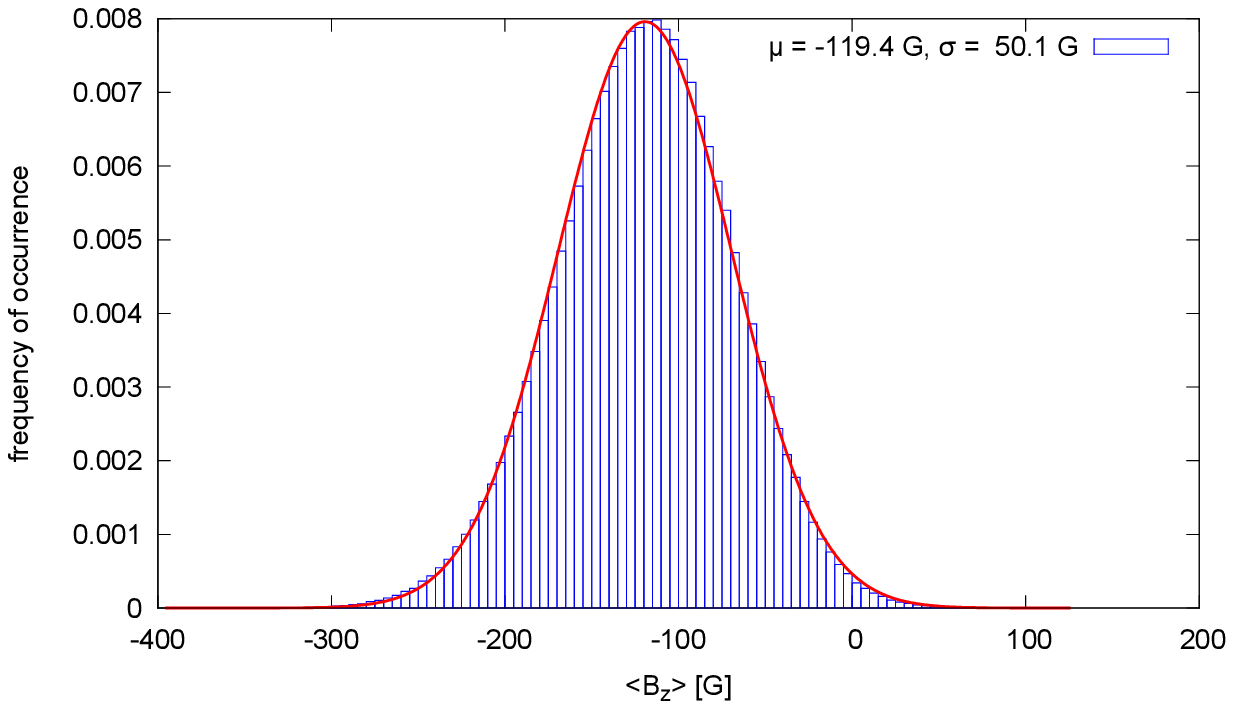}}
\caption{
Hen\,2-131: Regression detection of a $\left<B_{\rm z}\right>=-120\pm32$\,G 
mean longitudinal magnetic field with method R1, using the entire
spectrum (top), and corresponding distribution of 
$\left<B_{\rm z}\right>$ derived from method RM (bottom).}
\label{fig:hen2_131_b}
\end{figure}

\paragraph{\it Hen\,2-113:}
The central star of this bipolar PN is H-deficient, classified as 
a Wolf-Rayet star of type [WC11] (Mendez \& Niemela 
\cite{MendezNiemela1977}; Tylenda et al.\ \cite{Tylenda1993}).  
Hen\,2-113 was not yet observed by the \emph{Chandra} X-ray Observatory,
and no other information is available concerning the X-ray emission of this PN.

The target Hen\,2-113, with a rich emission line spectrum, 
shows a variable slope 
of the continuum in Stokes~$V$ spectra obtained at different epochs. Such 
variable continuum slopes were previously detected in our measurements 
of a few Herbig Ae stars with complex circumstellar environment, a few WR 
stars, and in the magnetic Of?p star HD\,148937, which is surrounded by the
circumstellar nebula NGC\,6164-65 that expands with a projected velocity of
about 30\,km/s (Leitherer \& Chavarria \cite{LeithererChavarria1987}).  The
Stokes~$V$ spectra for Hen\,2-113 were rectified before the magnetic
field measurements were made.

A weak, but significant mean longitudinal magnetic field of negative polarity 
was detected with both method R1 and RM on the first observing night: 
${\left<B_{\rm z}\right>=-58\pm18}$\,G and 
${\left<B_{\rm z}\right>=-58\pm24}$\,G, respectively, when using the entire
spectrum (wavelength set \emph{all}), and
${\left<B_{\rm z}\right>=-78\pm25}$\,G and 
${\left<B_{\rm z}\right>=-80\pm26}$\,G, respectively, when using the 
clean stellar lines only (wavelength set \emph{star}).
The latter detection is illustrated in 
the top panel of Fig.\,\ref{fig:hen2_113_b}. Note that the formal error
of the linear regression is particularly small in this case, mainly because
the spread of the data points in $x$ is large (cf. Eq.\,\ref{sigmaR1}) due 
to the presence of sharp emission lines (large $\mathrm{d}I/\mathrm{d}\lambda$).
As before, methods R1 and RM provide essentially the same error 
estimates, suggesting that the detection is solid.

In the case of Hen\,2-113, we have derived another independent error estimate
based on an experiment with simulated data (see Appendix~\ref{A2}).  This
experiment shows that, under somewhat idealized conditions, our analysis
method is expected to have a typical $1\,\sigma$ error of about $35$~G, given
the spectral resolution and signal-to-noise ratio of the present
observations. This error estimate is somewhat larger than those 
reported above from methods R1 and RM. We speculate that this mismatch may
be related to the fact that the number of lines in the synthetic spectrum is
lower than in the observed spectrum. Nevertheless, the numerical
experiment with synthetic spectra indicates that the errors obtained 
from the actual measurements in the range $\sigma_{\rm B} = 18$ to $30$~G 
are not unrealistic, and that the $3\,\sigma$ detection 
limit of our present measurements of Hen\,2-113 is as low as 
$|\left<B_{\rm z}\right>| \approx 120$~G.

\begin{figure*}
\mbox{\includegraphics[bb=28 28 580 360,width=12cm]{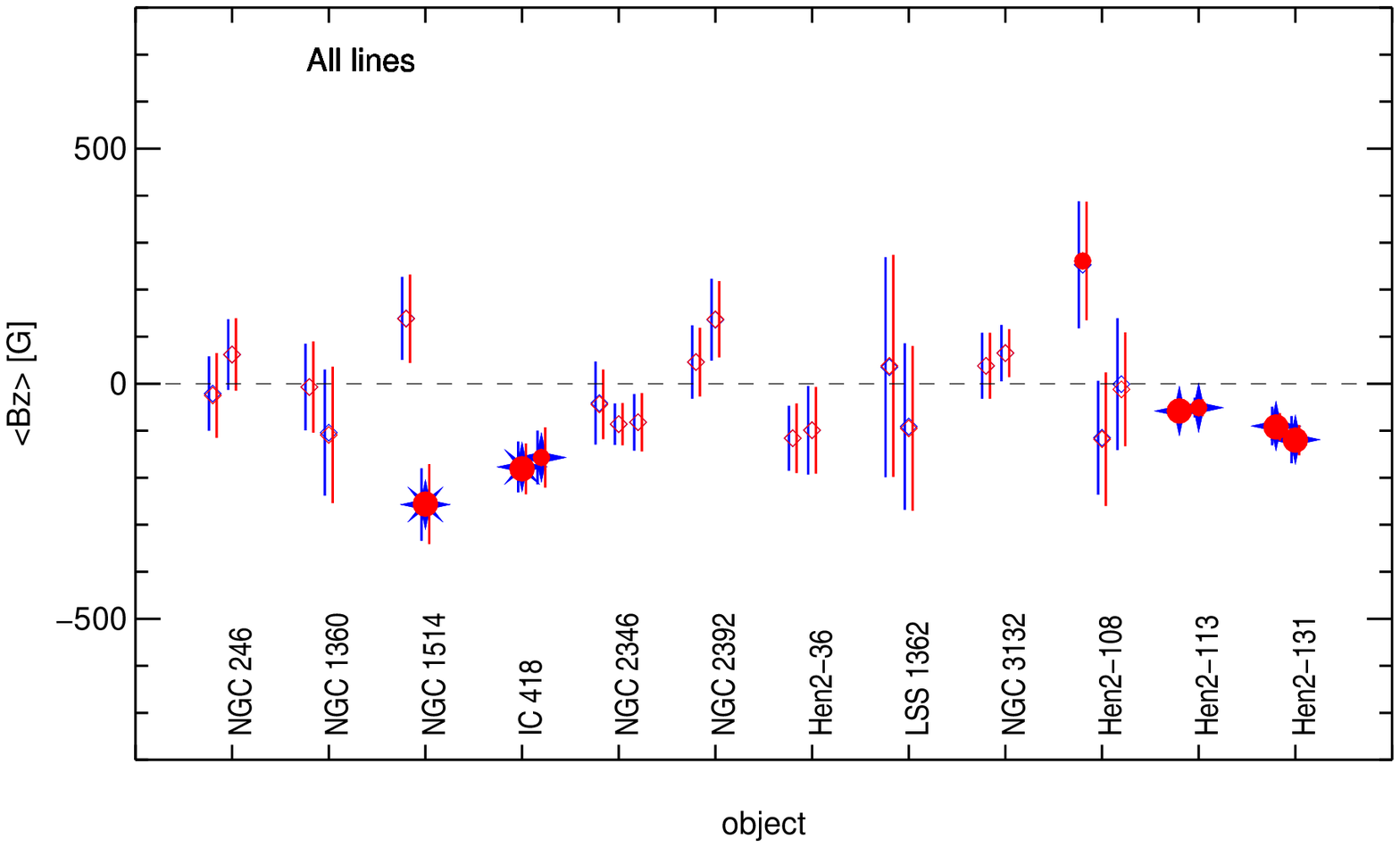}}\\
\sidecaption
\mbox{\includegraphics[bb=28 28 580 360,width=12cm]{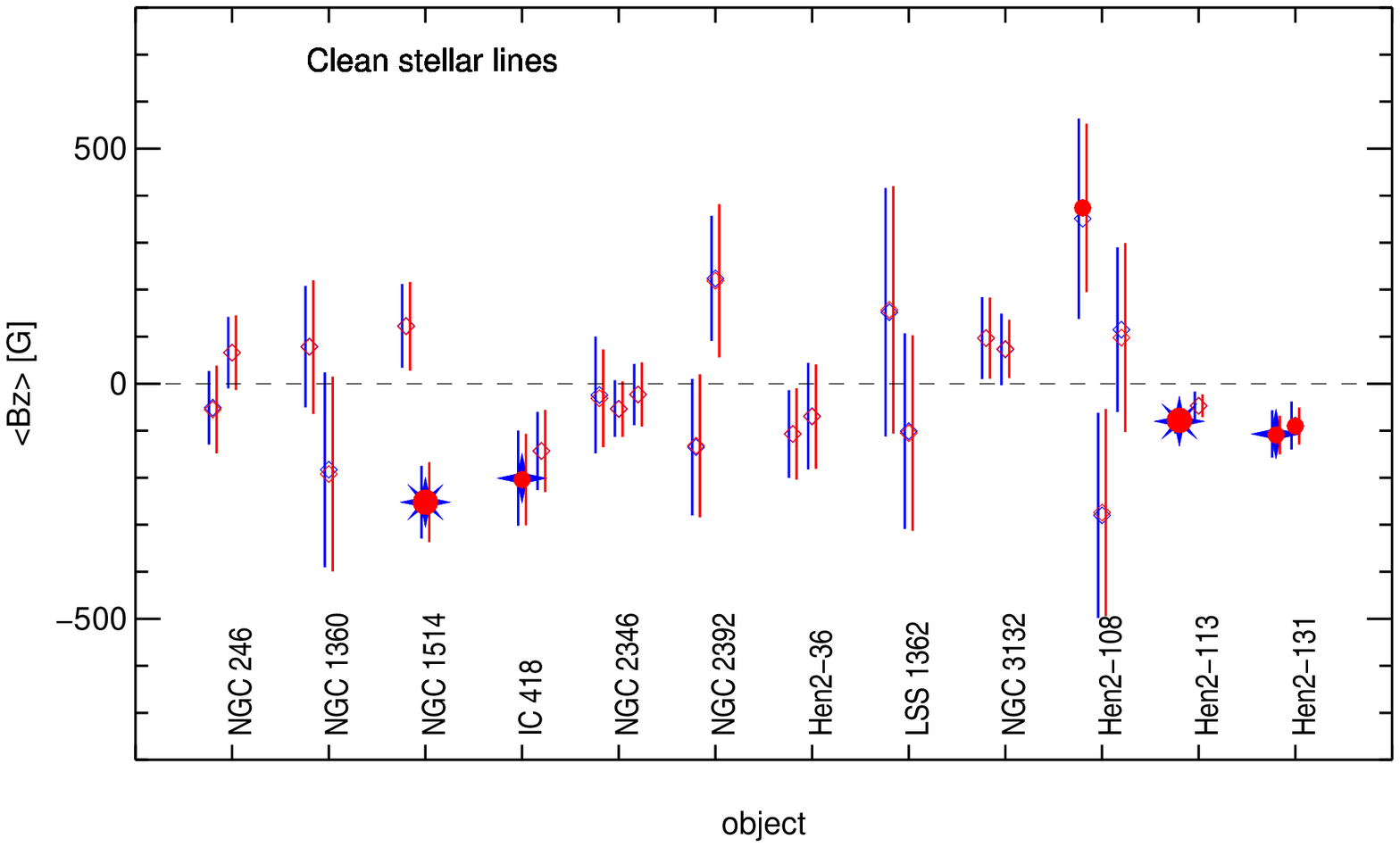}}
\caption{
Summary of our magnetic field measurements for the 12 targets of our
sample, using wavelength set \emph{all} (entire spectrum, 
top) and wavelength set \emph{star} (clean stellar lines only, bottom) as
listed in Table\,\ref{tab:log_meas}. Open symbols indicate non-detections, 
small and large filled dots (red) stand for $2\,\sigma$ and $3\,\sigma$ 
detections, respectively, with method R1, while (blue) stars mark $2\,\sigma$
(4 rays) and $3\,\sigma$ (8 rays) detections with method RM. Error bars
refer to $1\,\sigma$ errors from method R1 (red) and RM (blue), respectively.}
\label{fig:summary_all}
\end{figure*}

\paragraph{\it Hen\,2-131:}
According to Tylenda et al.\ (\cite{Tylenda1993}) the central star 
is a weak emission line star (WELS).  
No information is available concerning the X-ray emission of this elliptical
planetary nebula.

The Stokes~$I$ spectra appear to be variable. In
Fig.~\ref{fig:hen2_131_i} we present as an example the
\ion{C}{iii} $\lambda$\,4650.2 line profile variability
detected in the Stokes~$I$ spectra obtained on both observing nights.

Relying on method R1, and using the entire spectrum, a weak mean longitudinal 
magnetic field was detected on both observing nights at a formal significance
level of $\ga 3\,\sigma$: ${\left<B_{\rm z}\right>=-92\pm29}$\,G 
and ${\left<B_{\rm z}\right>=-120\pm32}$\,G, on the first and second night,
respectively. The latter detection is illustrated in 
Fig.\,\ref{fig:hen2_131_b} (top). Relying instead on method RM, 
we obtain very similar magnetic field strengths, but slightly larger
error bars: ${\left<B_{\rm z}\right>=-90\pm41}$\,G 
and ${\left<B_{\rm z}\right>=-119\pm50}$\,G
(Fig.\,\ref{fig:hen2_131_b}, bottom).
Similar results are derived when using the clean stellar lines
only, again indicating a mean longitudinal magnetic field of $\approx -100$~G,
but only with a significance of $2\,\sigma$.
Hen\,2-131 may thus be a marginal detection.

The $3\,\sigma$ upper limit for the mean longitudinal field of this
central star is as low as $|\left<B_{\rm z}\right>| \la 150$~G.

\section{Discussion}
\label{sect:discussion}

\subsection{Significance of our magnetic field measurements}
\label{sec:error_analysis}

Most of the stars of our sample were not studied at the achieved accuracy 
before, and the present study allows us to put further constraints 
on the strength of the magnetic fields in central stars of planetary nebulae. 
In Fig.\,\ref{fig:summary_all} we present a synopsis of the results of our
magnetic field measurements based on the entire spectrum (top) and on
clean stellar lines only (bottom).

The achievable measurement accuracy is limited by the comparatively low 
number of useful spectral lines for the magnetic field diagnostics in 
the central stars of planetary nebulae. In addition, since no linear 
polarization measurements exist for our sample stars, it is difficult 
to assess how our detections are affected by crosstalk. It is therefore
challenging to provide reliable estimates for $\left<B_{\rm z}\right>$.

In addition to the formal errors from the linear regression (method R1, 
Appendix~\ref{A11}), we have alternatively considered  an independent 
error estimate obtained from a bootstrapping analysis (method RM, 
Appendix~\ref{A1M}). As may be seen from Fig.\,\ref{fig:summary_all},
the errors from method RM turn out to be surprisingly similar to those 
from method R1. This close agreement suggests that our error estimates 
can be considered realistic.

\begin{table*}[htb]
\caption{
Physical parameters of the central stars of the PNe observed in the
present program (excluding those whose spectra are outshone
by an A-type companion). The quantity $|3\,\left<B_{\rm z}\right>|$ is the 
upper limit of the surface magnetic field derived from our measurements
and $B_{\rm s}^\ast$ and $B_{\rm s}^\dagger$ is the 
minimum magnetic field strength at the stellar surface that would be
required to shape the nebula according to Eq.\,(\ref{eq:Bs1}) and 
(\ref{eq:Bs2}), respectively.
}
\label{tab:cspn_dat}
\centering
\begin{tabular}{lrcccrrrrrrr}
\hline
\hline\noalign{\smallskip}
\multicolumn{1}{l}{Name}        &
\multicolumn{1}{c}{$T_{\rm eff}$} &
\multicolumn{1}{c}{$R_{\rm s}$}         &
\multicolumn{1}{c}{$\log (L/L_\odot)$}   &
\multicolumn{1}{c}{$d$}   &
\multicolumn{1}{c}{$\log \dot{M}$}   &
\multicolumn{1}{c}{$v_{\rm w}$}  &
\multicolumn{1}{c}{$v_{\rm rot}^{a)}$} &
\multicolumn{1}{c}{$|3\,\left<B_{\rm z}\right>|$}   &
\multicolumn{1}{c}{$B_{\rm s}^\ast$}   &
\multicolumn{1}{c}{$B_{\rm s}^\dagger$} &
\multicolumn{1}{c}{Ref.} \\
 &
\multicolumn{1}{c}{[kK]} &
\multicolumn{1}{c}{[$R_\odot$]} &
 &
\multicolumn{1}{c}{[kpc]} &
\multicolumn{1}{c}{[$M_\odot$/yr]} &
\multicolumn{1}{c}{[km/s]}   &
\multicolumn{1}{c}{[km/s]}   &
\multicolumn{1}{c}{[G]}      &
\multicolumn{1}{c}{[G]}      &
\multicolumn{1}{c}{[G]}      &
\multicolumn{1}{c}{cols.\,(2)--(8)} \\
\hline\noalign{\smallskip}
NGC\,246      & 150   & 0.20   & 4.20  & 0.50 &  $-$6.90 & 3500 & $>$77    &    $<~900$ &  1500~ & 
$2.3\times 10^5$ & 5,7,9   \\
NGC\,1360     &  97   & 0.30   & 3.64  & 0.93 & $-$10.60 & 1250 & $\sim$10 &    $<1350$ &  23~ &   1210~ & 2,3 \\ 
IC\,418       &  39   & 1.59   & 3.72  & 1.20 &  $-$7.50 &  700 & $>$56    &    $<~900$ &  12~ &     62~ & 1,8 \\
NGC\,2392     &  45   & 1.34   & 3.82  & 1.28 &  $-$8.10 &  300 & $\sim$10 &    $<1350$ &  11~ &     83~ & 1   \\
LSS\,1362     & 114   & 0.18   & 3.70  & 1.03 &  $-$8.70 & 2400 & $\sim$10 &    $<2400$ &  920~ &
$7.2\times 10^4$ & 2,3 \\
Hen\,2-108    &  34   & 2.60   & 3.92  & 5.80 &  $-$6.85 &  700 & $\sim$10 &    $<2100$ &  85~ &    320~ & 4   \\
Hen\,2-113    &  31   & 2.50   & 3.72  & 1.23 &  $-$6.10 &  160 & $\sim$10 &    $<~360$ &  23~ &     68~ & 6   \\ 
Hen\,2-131    &  32   & 3.50   & 4.07  & 3.30 &  $-$6.88 &  400 & $\sim$10 &    $<~450$ &  26~ &     85~ & 4   \\ 
\hline\noalign{\smallskip}
\end{tabular}
\tablefoot{
References: (1) Ruiz et al.\ (\cite{Ruiz2013}); 
(2) Traulsen at al. (\cite{Traulsen2005}); 
(3) Herald \& Bianchi (\cite{Herald2011});
(4) Kudritzki et al.\ (\cite{Kudritzki2006});
(5) Koesterke et al.\ (\cite{Koesterke1998});
(6) De Marco \& Crowther (\cite{DeMarco1998});
(7) Bond \& Ciardullo (\cite{BondCiardullo1999});
(8) Prinja et al.\ (\cite{Prinja2012});
(9) Rauch et al.\ (\cite{Rauch2003}).
$^{a)}$\,Assuming $v_{\rm rot}=10$~km/s where no measurements are available.}
\end{table*}

If we require that a detection is only valid if it is found with both
method R1 and method RM when using the wavelength set \emph{star}
(clean stellar lines only), then we count two 
$3\,\sigma$ detections, namely NGC\,1514 (second night) and Hen\,2-113
(first night), and two cases where our measurements indicate a magnetic field
at the $2\,\sigma$ significance level, namely IC\,418 (second night)
and Hen\,2-131 (first night). IC\,418 is, in fact, a $3\,\sigma$ detection
with wavelength set \emph{all}. For the two emission line objects 
\mbox{Hen\,2-113} and \mbox{Hen\,2-131}, the formal error is remarkably small, 
in the case of \mbox{Hen\,2-113} even smaller than the theoretical estimate 
derived from numerical experiments with synthetic spectra (Appendix~\ref{A2}),
and hence these formal detections must be considered with some caution.

In all cases, the magnetic field is detected only during one out of two 
useful observing nights, which, in view of the expected 
rotation periods of only a few days, might be attributed to rotational
modulation by large-scale magnetic structures.

In summary, we cannot claim a clear positive detection of a magnetic field 
in any of the PN central stars of our sample. What we can do, though, 
is to estimate individual upper limits of $|\left<B_{\rm z}\right>|$, 
as set by the quality of our measurements and the sensitivity of our 
analysis method (see Table\,\ref{tab:cspn_dat}, Col.\,9). In the following,
we investigate whether such upper limits are sufficient to rule out the PN
magnetic shaping scenario.

\subsection{The magnetized stellar wind}
\label{sec:magshape1}
If large-scale (dipole-like) magnetic fields are present at the surface 
of a central star, they are carried along with the fast stellar wind and 
are wound up due to stellar rotation. At large distances from the star, the 
field is essentially toroidal (i.e. has only an azimuthal
component). The impact of the magnetic field on the shaping of the 
shocked wind and the inner nebula is described by the magnetization 
parameter $\sigma_{\rm M}$, the ratio of magnetic to kinetic energy density of 
the magnetized stellar wind (see, e.g., Chevalier \& Luo 
\cite{ChevalierLuo1994}):

\begin{eqnarray}
\sigma_{\rm M} &=& \frac{B^2}{4\,\pi\,\rho\,v_{\rm w}^2} = 
           \frac{B_{\rm s}^2\,R_{\rm s}^2}{\dot{M}\,v_{\rm w}}\,\left(\frac{v_{\rm
               rot}}{v_{\rm w}}\right)^2 \nonumber \\
       &=& 7.7\times 10^{-5}
            \frac{\left(\frac{B_{\rm s}}{1\,\mathrm{G}}\right)^2\,
            \left(\frac{R_{\rm s}}{\mathrm{R}_\odot}\right)^2\,
            \left(\frac{v_{\rm rot}}{v_{\rm w}}\right)^2} 
            {\left(\frac{\dot{M}}{10^{-8}\,
            \mathrm{M}_\odot/\mathrm{yr}}\right)\,
            \left(\frac{v_{\rm w}}{1000\,\mathrm{km/s}}\right)}\, ,
\end{eqnarray}
where $B_{\rm s}$ is the magnetic field at the stellar surface, $R_{\rm s}$ the
stellar radius, and $\dot{M}$ and $v_{\rm w}$ the mass loss rate and the
terminal velocity of the central star's fast wind. Theoretical
considerations show that significant deviation from spherical nebular
expansion requires a minimum magnetization of the stellar wind 
given by  $\sigma_{\rm M} \ga 10^{-4}$ (Chevalier \& Luo \cite{ChevalierLuo1994}). 
While the wind is essentially unaffected by the magnetic field
at such low values of $\sigma_{\rm M}$, the magnetic pressure becomes
significant inside the hot bubble of shocked gas, thus allowing for magnetic 
shaping of the nebula (see Sect.\,\ref{sec:magshape2}).

The condition $\sigma_{\rm M} \ga 10^{-4}$ can be expressed as 
\begin{eqnarray}
B_{\rm s} \ga B_{\rm s}^\ast \approx 
\frac{\left(\dot{M}/10^{-8}\,\mathrm{M}_\odot/\mathrm{yr}\right)^{1/2}\,
\left(v_{\rm w}/1000\,\mathrm{km/s}\right)^{1/2}}
             {\left({R_{\rm s}}/\mathrm{R}_\odot\right)\,
              \left({v_{\rm rot}}/{v_{\rm w}}\right)}\, ,
\label{eq:Bs1}
\end{eqnarray}
where $B_{\rm s}^\ast$ is in units of Gauss.

We compiled the relevant physical parameters of our program stars
in Table\,\ref{tab:cspn_dat} (except for the four objects whose
spectra are dominated by an A-type companion), allowing us to  estimate 
$B_{\rm s}^\ast$ according to Eq.\,(\ref{eq:Bs1}), assuming $v_{\rm rot} 
\approx v\,\sin\, i$, or  $v_{\rm rot} \approx 10$~km/s where no measurement
of $v\,\sin\,i$ is available. For most of the central 
stars of our sample, $B_{\rm s}^\ast$ turns out to be weaker than $100$~G 
(see Table\,\ref{tab:cspn_dat}). Note that for the central 
stars of IC\,418 and  NGC\,2392, $B_{\rm s}^\ast$ is particularly low because 
of the high rotational velocity and the low wind speed, respectively.

Assuming that any potential
large-scale magnetic field at the surface of the central stars would be
dipole-like, we make the statistical approximation 
$B_{\rm s} \approx 3\,\left<B_{\rm z}\right>$. 
The upper limit of $B_{\rm s}$ derived from out measurements, 
$|3\,\left<B_{\rm z}\right>|$, is listed individually for each target
in Table\,\ref{tab:cspn_dat} (Col.\,9).

We find that, except for two 
of our targets (NGC\,246 and LSS\,1362), the upper limit for 
$\left<B_{\rm z}\right>$ that excludes magnetic shaping lies at least 
a factor of $10$ below the 
upper limit provided by our measurements. For most of our targets, we 
therefore cannot rule out the presence of a stellar surface magnetic field 
that is strong enough for shaping the present fast wind of the 
central stars.

We have to keep in mind, however, that the shaping of the PN must have 
occurred during a preceeding period of time, starting in the early stages 
of post-AGB evolution.
We are therefore interested to know whether the magnetic field was
strong enough to influence the stellar wind when the nebula
was just about to form. For this purpose, we extrapolate $B_{\rm s}$ 
backward in time, starting at the current value, and assuming that (i) the 
stellar luminosity is constant, $T_{\rm eff}^4\,R_{\rm s}^2$ = const.; 
(ii) the angular momentum of the star is conserved, $v_{\rm rot}\,R_{\rm s}$ 
= const.; (iii) the fast wind velocity scales with the escape velocity,
$v_{\rm w}^2\,R_{\rm s}$ = const.; (iv) $\dot{M}$ = const.; and (v) the total 
magnetic flux is conserved, $B_{\rm s}\,R_{\rm s}^2$ = const., implying that 
$\sigma_{\rm M}$ increases strongly during the time evolution as 
$\sigma_{\rm M} \sim T_{\rm eff}^5$. 

To ensure that the magnetization parameter $\sigma_{\rm M}$ exceeds the critical 
value of $10^{-4}$ already at an earlier time, $t=t_0$, the magnetic field 
strength at the current time, $t=t_1$, must fulfill the condition
\begin{eqnarray}
B_{\rm s} \ga B_{\rm s}^\dagger=B_{\rm s}^\ast\,
\left(\frac{T_{\rm eff}(t_1)}{T_{\rm eff}(t_0)}\right)^{5/2}\, .
\label{eq:Bs2}
\end{eqnarray}

In Table\,\ref{tab:cspn_dat}, we also provide the value of $B_{\rm s}^\dagger$
for the assumption $T_{\rm eff}(t_0)=20$~kK.
We see that $B_{\rm s}^\dagger$ clearly exceeds the upper limit for the 
measured stellar field strength, $B_{\rm s} \approx 3\,\left<B_{\rm z}\right>$, 
in the case of NGC\,246 and LSS\,1362, indicating that magnetic shaping
is not expected to play any role for these two round / elliptical objects. 
On the other hand, we find that the upper limit for $B_{\rm s}$ estimated 
from our magnetic field measurements is still about a factor of $5$ to $10$ 
higher than $B_{\rm s}^\dagger$ for the five PNe IC\,418, NGC\,2392,
Hen\,2-108, Hen\,2-113, and Hen\,2-131. For these objects, our measurements do
not rule out the possibility that their nebulae have been subject to magnetic
shaping since early on in their post-AGB evolution when $T_{\rm eff} \approx
20$~kK, or even since shortly after the end of the AGB evolution (see below).
The situation in less clear for NGC\,1360 where $B_{\rm s}^\dagger$ is
intermediate between the two extremes mentioned above.

The estimation of the magnetization parameter at even earlier times,
near the tip of the AGB ($t=t_{\rm AGB}$, $T_{\rm eff} \approx 5$~kK), 
is more uncertain because 
the above assumption of constant mass loss rate is no longer valid 
below $T_{\rm eff} \approx 20$~kK ($t < t_0$).
Assuming instead that $\dot{M}$ decreases by a factor of $100$ between 
$t=t_{\rm AGB}$ and $t=t_0$, while at the same time $v_{\rm w}$
increases by a factor of $100$, we find that, roughly,
 $\sigma_{\rm M}(t_{\rm AGB})/\sigma_{\rm M}(t_0) \approx 0.1$, in agreement with the
numbers quoted by Garc{\'{\i}}a-Segura et al.\ (\cite{GarciaSegura1999})
based on Reid at al.\ (\cite{Reid1979}). 
We conclude that the slow but massive AGB wind is probably unaffected 
by magnetic shaping as long as $B_{\rm s} < 10 B_{\rm s}^\dagger$, which is
the case for most of our targets. For only two targets, IC\,418, and 
NGC\,2392, our measurements do not even rule out a marginal magnetic 
influence on the very early post-AGB wind.

\subsection{Magnetic shaping of the hot bubble}
\label{sec:magshape2}

The free wind of the central star is bounded by a strong reverse shock
at radius $R_1$ where the kinetic energy of the wind is thermalized. The
shocked wind fills the so-called hot bubble, bounded by the contact
discontinuity at outer radius $R_2$.
The strength of the magnetic field in the immediate pre-shock region, 
$B_{\rm p}$, may be estimated from the stellar surface magnetic field, 
$B_{\rm s}$, as 
$B_{\rm p} \sim B_{\rm s}\times\,v_{\rm rot}/v_{\rm w}\times\,R_{\rm s}/R_1$
(e.g., Chevalier \& Luo \cite{ChevalierLuo1994}), where 
$v_{\rm rot}$ and $v_{\rm w}$ are the (equatorial) rotation velocity 
of the central star and the outflow velocity of its fast wind, while 
$R_{\rm s}$ and $R_1$ is the radius of the central star and the inner radius
of the hot bubble, respectively. 
Assuming $R_1 = 0.01$~pc, we evaluated the ratio $B_{\rm p}/B_{\rm s}$
for the central stars of our target list. The ratio lies in the range
$1.7\times 10^{-9}$ to $3.5\times 10^{-7}$. We thus conclude that the strength
of magnetic fields in the pre-shock region cannot exceed a few hundred $\mu$G.
Such fields are too weak to influence the structure of the strong wind shock,
however, according to the Chevalier \& Luo (\cite{ChevalierLuo1994}) model,
the magnetic field builds up downstream as a consequence of the 
compressive flow inside the hot bubble.

The structure of the magnetized hot bubble is illustrated in Fig.\,1 of
Chevalier (\cite{Chevalier1992}) for the quasi-spherical case, assuming
$\sigma_{\rm M} = 0.001$, which is readily applicable to the shocked wind
bubbles of PNe. Across the shock, $B$ jumps from the pre-shock value
$B_{\rm p}$ to a post-shock value of $B_{\rm X}(R_1) \approx 16\times B_{\rm  p}$.
Because of the negative velocity gradient inside the hot bubble, the magnetic 
flux density increases outwards and rises to $B_{\rm X}(R_2) \approx 320\times
B_{\rm p}$ at the outer edge of the hot bubble, assuming that $R_2 \approx
5\times R_1$.

For sufficiently large $\sigma_{\rm M}$, the magnetic field has a direct 
dynamical effect on the expansion of the hot bubble. Figure\,1 
of Chevalier (\cite{Chevalier1992}) shows that 
the magnetic pressure, $B_{\rm X}^2/8\pi$, can even exceed the thermal gas
pressure, $n_{\rm e}\,k\,T$, in the outer parts of large (old) bubbles.  At
the same time, the magnetic tension gives rise to a non-spherical expansion by
reducing the driving force in the equatorial direction, thus leading to an
elongated shape in the polar direction. This mechanism is at the heart of the
magnetic shaping scenario.

According to the model assumptions made above in
Sect.\,\ref{sec:magshape1}, the hot bubble of older nebulae are
expected to harbor stronger magnetic fields than younger nebulae. For constant
$R_1$, we find that $B_{\rm p}$ increases as $ B_{\rm p} \sim T_{\rm eff}^3$.
Since $T_{\rm eff}$ typically increases by a factor of $10$ during the
post-AGB evolution, $ B_{\rm p}$ would increase roughly by a factor of $1000$.
Taking into account that at the same time the central cavity expands by
roughly a factor of $5$ (see, e.g., Fig.\,5 of Sch\"onberner et
al.\ \cite{Schoenberner2005}), we estimate that $B_{\rm p}$ is about $200$
times larger in old compared to young PNe.

\subsection{X-ray emission}
\label{sec:xray}
According to the order of magnitude estimates given in 
Sect.\,\ref{sec:magshape2}, we may assume, as an upper limit, that the 
magnetic field varies in the X-ray emitting hot bubble between a few 
mG near the inner shock up to $100$\,mG strength near 
the  outer contact discontinuity.
Weak magnetic fields of this order of magnitude can easily suppress thermal 
conduction in the direction perpendicular to their field lines. According to
Spitzer (\cite{Spitzer1962}), the factor by which the thermal conductivity 
is reduced in this direction with respect to the conductivity along the
field lines (or the field-free case) can be expressed for a pure hydrogen
plasma as
\begin{eqnarray}
\frac{D_\bot}{D_\|} \approx 7.6\times 10^{-16}\,
\frac{n_{\rm e}^2}{T_6^3\,B_{\mu\mathrm{G}}^2}\, ,
\label{eq:kratio}
\end{eqnarray}
where $n_{\rm e}$ is the electron density in [cm$^{-3}$], $T_6$ the
electron temperature in [$10^6$~K], and $B_{\mu\mathrm{G}}$ the magnetic
flux density in units of [$10^{-6}$~G] (see also Balbus\ \cite{Balbus1986}). 
In the X-ray emitting cavity ($n_{\rm e} \approx 10$, $T_6 \ga 10$,  
$B_{\mu\mathrm{G}} \ga 1$) we obtain  ${D_\bot}/{D_\|} \la 10^{-13}$, 
indicating that any thermal conduction perpendicular to the magnetic 
field lines is expected to be very effectively suppressed, even for field
strengths that are many orders of magnitude below $1 \mu$G.  

If the magnetic field has a purely toroidal orientation in the X-ray
emitting central cavity, thermal conduction in the radial direction
should be completely suppressed, even if the field is much weaker than
those deduced in our present study. As a consequence, the luminosity of the
diffuse X-ray emission should be significantly lower, and the characteristic
X-ray temperature should be significantly higher than what is typically 
found in X-ray observations (see, e.g., Steffen et al.\ \cite{Steffen2008}).
This dilemma may indicate that the nebular field geometry is different from
purely toroidal. Alternatively, other physical mechanisms, such as turbulent
mixing due to hydrodynamic instabilities at the contact discontinuity 
(e.g., Stute \& Sahai \cite{StuteSahai2006}; Toala \& Arthur\ 
\cite{Toala2014}), may provide an efficient channel for radial heat 
exchange.

\subsection{Central star magnetic field and nebular morphology}
Two out of the three bipolar nebulae of our sample harbor a central
star with an A-type binary companion (NGC\,2346 and Hen\,2-36). In
these cases, the magnetic field measurement of the central stars
proper is impossible due to contamination of their spectrum by the
A-type companion. For the central star of the remaining bipolar
PN, Hen\,2-113, our magnetic field measurements provide
marginal support for a magnetic origin of the bipolar nebular 
structure.

For the central star of the elliptical PN LSS\,1362, our analysis
showed that magnetic shaping is expected to be unlikely
and is at best restricted to the present evolutionary stage. This might
indicate that the elliptical shape of LSS\,1362 has a non-magnetic 
background, but the constraints are weak. For the remaining six
elliptical PNe, the magnetic shaping hypothesis cannot be constrained
by our magnetic field estimates.

The only case where we can rule out significant magnetic shaping is
NGC\,246. Since this PN is indeed round, there is no reason here
to invoke a non-magnetic shaping mechanism.
No constraints can be derived for the other round PN, Hen\,2-108.

In view of the poor statistics, it is clearly impossible to establish any firm
empirical relation between the strength of the central star's magnetic field
and the morphological type of the associated planetary nebula.
Given the rather low measurement accuracy, we also cannot find any clear
inconsistencies that would invalidate the magnetic shaping hypothesis.

\subsection{Central star magnetic field and X-ray emission}
According to Sect.\,\ref{sec:xray}, both weak and strong magnetic fields 
suppress thermal conduction with essentially the same efficiency. From
this perspective, we would not expect any correlation between the strength 
of the stellar magnetic field and the luminosity in {diffuse X-rays}. 
However, the thermal structure and density profile of the hot bubble
can be significantly altered by the presence of sufficiently strong magnetic
fields (cf.\ Sect.\,\ref{sec:magshape2}).
Obviously, an empirical verification of a possible relation between 
stellar magnetic field strength and diffuse X-ray emission is
impossible with our sample: only two of the elliptical nebulae 
show diffuse X-ray emission (IC\,418 and NGC\,2392) and in both cases 
only a (rather high) upper limit for the stellar magnetic field could
be deduced.

On the other hand, the same two objects suggest that the presence of a central
{point source of hard X-ray emission} is not related to the strength of
the stellar magnetic field either. Our measurements indicate that
the stellar magnetic field is unlikely to be significantly stronger in 
NGC\,2392  than in IC\,418, and yet NGC\,2392 harbors an X-ray point source 
while IC\,418 does not, which seems difficult to reconcile with the idea 
of a magnetic origin of the hard X-ray emission.

\section{Conclusions}
\label{sect:conclusions}
We have analyzed spectropolarimetric observations of 12 central stars
of planetary nebulae carried out at the European Southern Observatory with
FORS 2 mounted on the 8-m Antu telescope of the VLT.
We find marginal evidence for weak mean longitudinal magnetic fields of 
about $200$\,G in the central star of the young elliptical nebula IC\,418,
and for even weaker magnetic fields of about 100\,G in the emission-line 
spectra of the Wolf-Rayet type central star Hen\,2-113 and the weak emission 
line star Hen\,2-131. In general, however, we can only estimate upper 
limits for the  mean longitudinal magnetic fields $|\left<B_{\rm z}\right>|$ 
in the range $120$ to $800$~G. We conclude that strong magnetic fields 
exceeding $1$~kG are not widespread among PNe central stars. 

Some of the observed central stars, including IC\,418 and Hen\,2-131, 
show a significant night-to-night spectrum variability of the stellar line
profiles, which may be attributed to rotational modulation due to magnetic
features or inhomogeneities in the stellar wind. Follow-up monitoring of 
these targets may be worthwhile to confirm the rotational/magnetic variability
and to determine the related timescales.

Interestingly, the hypothesis that planetary nebulae are shaped by magnetic 
fields is not ruled out by our measurements. Theoretical order of magnitude
estimates suggest that even weak magnetic fields well below the detection 
limit of our measurements are sufficient to contribute to the shaping of 
the surrounding nebulae throughout their evolution. We have to conclude that,
at the present accuracy, the available measurements of central star magnetic 
fields can neither support nor rule out the magnetic shaping hypothesis for
most of our targets. We can only infer for NGC\,246 and LSS\,1362 that
magnetic shaping is not playing any significant role.

The only clear $3\,\sigma$ detection of a 250 G mean 
longitudinal field is achieved for the A-type companion of the central 
star of NGC\,1514. Even though the mass loss of an A-type star is negligible 
compared to that of the true central star, the magnetic field of the A-type 
companion might interact with the wind of the central star. Whether or not
such interaction would play any role in shaping the surrounding 
planetary nebula is unclear. In any case, we can assume that both stars 
formed and evolved in a common environment. The presence of a magnetic field
in the companion of the central star is of great interest in the context of 
understanding the origin of magnetic fields in A-type stars, since only a
few magnetic Ap stars are known to belong to close binary systems.


\begin{acknowledgements}
C.S.\ was supported by funds of DFG (project SCHO 394/29-1) and Land 
Brandenburg (SAW funds from WGL), and also by funds of PTDESY-05A12BA1.
We thank Martin Wendt for fruitful discussions of statistical methods,
and the referee, John Landstreet, for constructive criticism 
that led to significant improvement of this paper.

\end{acknowledgements}

\newpage
------------------------------------------------------------------------
\begin{appendix}
\section{Selected wavelength regions}
\label{A0}

\begin{figure*}
\centering
\includegraphics[angle=0,width=0.94\textwidth]{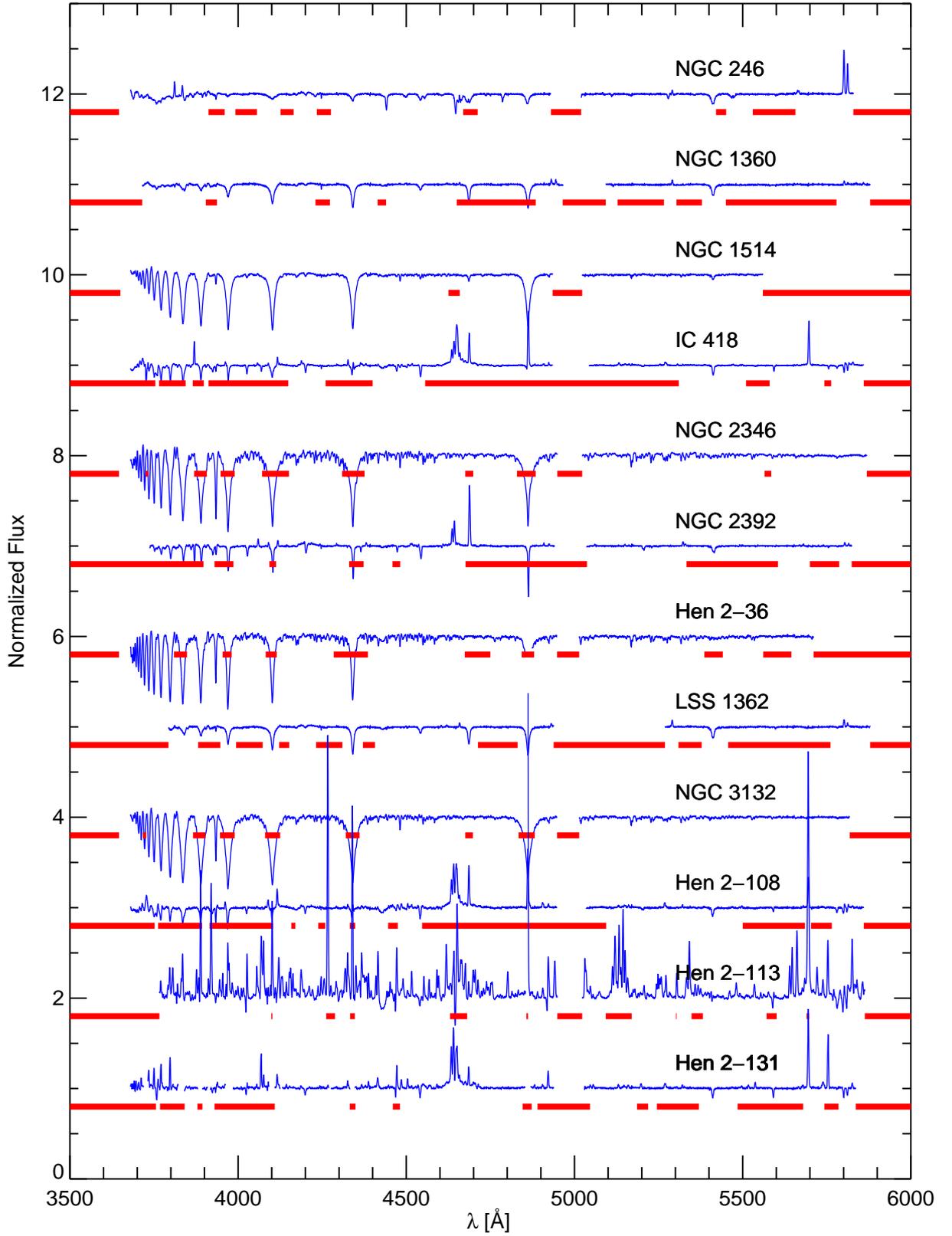}
\caption{
Normalized FORS\,2 Stokes~$I$ spectra of the $12$ central 
stars in our sample, displayed with a vertical offset of 1 unit between 
adjacent spectra. For each object, the displayed spectra (blue) define 
wavelength set \emph{all} (entire spectrum), gaps indicating the excluded 
regions, while wavelength set \emph{star} (clean stellar lines only) is 
obtained by excluding the regions marked by the (red) horizontal bars $0.2$ 
units below the respective stellar continuum.}
\label{fig:regions}
\end{figure*}

\paragraph{Wavelength region \emph{all}} comprises the entire spectral
range, but excludes the nebular [\ion{O}{iii}] emission lines near 
$5000$\,\AA\ and pixels affected by instrumental artifacts, as identified 
in the null spectra and removed by $3\,\sigma$ clipping. For each object, 
the spectra shown in Fig.\,\ref{fig:regions} are plotted over the useful 
spectral range defined in this way (referred to as wavelength set 
\emph{all} in this work).

\paragraph{Wavelength region \emph{star}} is obtained from wavelength
region \emph{all} by excluding additional spectral windows that contain
nebular emission lines. The latter could be identified in the original 
CCD frames in traces located above and below the spectrum of the central
star. The spectral regions excluded in this procedure are indicted as
horizontal bars 0.2 units below the respective stellar continuum of each
object in Fig.\,\ref{fig:regions}.

\section{Deriving $\left<B_{\rm z}\right>$ and $\sigma_B$ 
from linear regression}
\label{A1}
The mean longitudinal magnetic field $\left< B_{\rm z}\right>$ is 
derived from the fundamental relation
\begin{eqnarray} 
y(\lambda) \equiv \frac{V}{I} = 
-\frac{g_{\rm eff}\, e \,\lambda^2}{4\pi\,m_{\rm e}\,c^2}\,
\frac{1}{I}\,\frac{{\rm d}I}{{\rm d}\lambda} \left<B_{\rm z}\right>
\equiv x(\lambda)\,\left<B_{\rm z}\right>\, 
\label{eqn:one}
\end{eqnarray} 
(cf.\ Eq.\,(\ref{eqn:vi}) and related text for the definition of the 
different symbols). For the present analysis, we use the original wavelength 
scale provided by our pipeline ($\Delta\lambda = 0.75$\,\AA), avoiding 
interpolation to finer wavelength steps.

\subsection{Method R1}
\label{A11}
Given the original dataset $\left\{x_i(\lambda), y_i(\lambda)\right\}_{i=1,N}$, 
where $N$ is the total number of considered spectral bins, and assuming 
that relation (\ref{eqn:one}) is valid both for absorption and emission
lines, with $g_{\rm eff}\approx 1.2$, we compute $\left<B_{\rm z}\right>$ from 
linear regression (e.g., Press et al.\ \cite{Press2007}) as:
\begin{equation}
\label{eqn:A2}
\left<B_{\rm z}\right> = \frac{\overline{x\,y} -\overline{x}\,\overline{y}}
{\overline{x^2} - \overline{x}^2},
\end{equation}
where we have defined
\begin{eqnarray}
\overline{x} &=& \frac{\sum_N w_i\,x_i}{\sum_N w_i},\;\;
\overline{x^2} =\frac{\sum_N w_i\,x_i^2}{\sum_N w_i},\;\; \nonumber \\
\overline{y} &=& \frac{\sum_N w_i\,y_i}{\sum_N w_i},\;\;
\overline{x\,y} =\frac{\sum_N w_i\,x_i\,y_i}{\sum_N w_i}\, .
\end{eqnarray}
The weight of each pixel is given by the square of the signal-to-noise
ratio of Stokes $V$, \mbox{$w_i=(S/N)_i^2$.} Considering the propagation of 
errors, assuming that the errors in $x$ are negligible,
we obtain the formal 1\,$\sigma$ uncertainty of $\left<B_{\rm z}\right>$ as
\begin{eqnarray}
\label{sigmaR1}
\sigma_{B} &=& \sqrt{
\frac{1}{\sum_N w_i}\,\frac{1}{\overline{x^2} - \overline{x}^2}} \nonumber \\
&=& \frac{1}{\sqrt{N}}\,\frac{1}{(S/N)_{\rm rms}}\,
    \frac{1}{\sqrt{\overline{x^2} - \overline{x}^2}}.
\end{eqnarray}
As a sanity check, we also compute the quantity
\begin{eqnarray}
\label{chi2}
\frac{\chi_{\rm min}^2}{\nu} &=& \frac{1}{N-2}\,\sum_N w_i(y_i-f_i)^2\, ,
\end{eqnarray}
where $f_i$ is the value obtained from the best fitting straight line,
\begin{eqnarray}
\label{fitfun}
f_i = f_0 + \left<B_{\rm z}\right>\,x_i ,
\end{eqnarray}
 with 
\begin{eqnarray}
\label{f0}
f_0 = \frac{\overline{x^2}\,\overline{y} -\overline{x}\,\overline{x\,y}}
{\overline{x^2} - \overline{x}^2}.
\end{eqnarray}
Whenever $\chi_{\rm min}^2/\nu > 1$, $\sigma_{B}$ is multiplied by the factor
$\sqrt{\chi_{\rm min}^2/\nu}$ to obtain the final error estimate of
$\left<B_{\rm z}\right>$. In general, this factor is $\la 1$, and
never exceeds $1.07$ for the applications considered in this paper.

\subsection{Method RM}
\label{A1M}
In this Monte-Carlo type approach, we generate $M=10^6$ statistical
variations of the original dataset 
$\left\{x_i(\lambda), y_i(\lambda)\right\}_{i=1,N}$
by bootstrapping (e.g., Rivinius et al.\  \cite{Rivinius2010}).
Each of the $M$ artificial datasets comprises $N$ data points, 
obtained by randomly drawing $N$ times a data point from the original 
sample, assigning the same probability of being drawn to all data points.
For each of these $M$ generated datasets, we derive the value 
$\left<B_{\rm z}\right>_m$ using Eq.\,(\ref{eqn:A2}), taking the
weights $w_i$ of the individual data points into account.
The expectation value of $\left<B_{\rm z}\right>$ is given by the 
mean of the distribution, 
\begin{equation}
\label{BzRM}
\overline{\left<B_{\rm z}\right>} = \frac{1}{M}\,\sum_{m=1}^M \left<B_{\rm z}\right>_m\, ,
\end{equation}
and the $1\,\sigma$ error is estimated from
the standard deviation of the distribution, even if the latter
is not necessarily Gaussian:
\begin{equation}
\label{sigmaRM}
\sigma_{\overline{B}} = \sqrt{
\frac{1}{M}\,\sum_{m=1}^M \left(\left<B_{\rm z}\right>_m - 
                        \overline{\left<B_{\rm z}\right>}\right)^2}\, .
\end{equation}
In this scheme, outliers in the original dataset will automatically translate
into an increased error estimate, even if the formal error ($1/\sqrt{w_i}$) 
assigned to such data points is small.

\section{Error of $\left<B_{\rm z}\right>$ from simulated data}
\label{A2}

Yet another independent estimate of the error margin of our magnetic field 
measurements is obtained from the following test based on simulated data. 
Among our targets, we select the [WC]-type central star \mbox{Hen2-113}. 
With the Potsdam
Wolf-Rayet (PoWR) model atmosphere code (Gr\"afener et al.\ 
\cite{Graefener2002}; Hamann \& Gr\"afener \cite{Hamann2003}) 
we calculate a synthetic spectrum 
that is similar to the observed spectrum of this star. 
We restrict the normalized synthetic 
spectrum to the observed wavelength range (4000 -- 6000\,\AA),
convolve it with a Gaussian of 3\,\AA\ according to  the spectral
resolution of FORS\,2, and bin the data to the pixel size of 0.66\,\AA. 

Then we add to the simulated spectrum artificial noise as a Gaussian
distribution with a standard deviation equal to $SN^{-1}$. The signal-to-noise
ratio, $SN$,  is set to 920. This corresponds to the quality of our
\mbox{Hen2-113} observations after all spectra from one
night that belong to the same polarization channel have been coadded.   

\begin{figure}
\centering
\mbox{\includegraphics[angle=-90,width=0.46\textwidth]{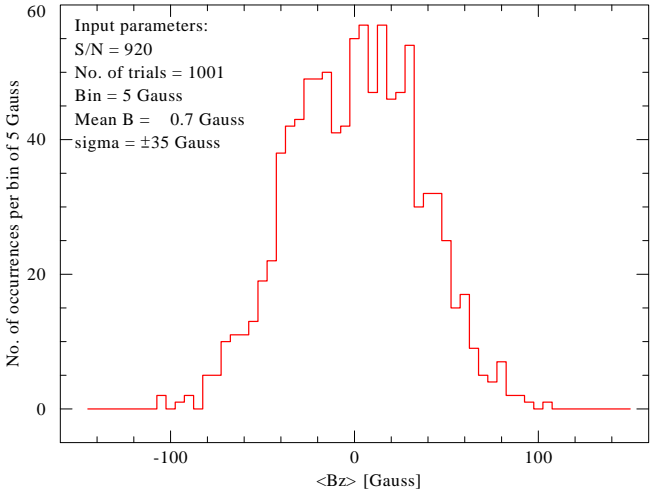}}
\caption{Distribution of 1001 $\left<B_{\rm z}\right>$ measurements on 
simulated data for \mbox{Hen2-113} with input field $\left<B_{\rm z}\right>=0$ 
and artificial noise corresponding to S/N = 920 per pixel in each channel.
As the results show, the method has in this case a $1\,\sigma$ error
of 35\,Gauss.}
\label{fig:simB}
\end{figure}

Two such simulated spectra with independent statistical noise are now
taken to mimic the ordinary and the extra-ordinary polarization
channel, respectively. These data are analyzed for their Zeeman shift in
the same way as the real observations. The result is a
``measured'' field strength $\left<B_{\rm z}\right>$. 
The whole procedure is now repeated 1001 times with the same simulated
data, but independent artificial noise. The distribution of the obtained
field strength $\left<B_{\rm z}\right>$ is plotted as a histogram in
Fig.\,\ref{fig:simB}. The obtained $\left<B_{\rm z}\right>$ values 
scatter around zero with a mean deviation of 35\,Gauss. 

Summarizing, this test was based on the following assumptions: (1) the
line spectrum is similar to our simulated, normalized  spectrum for 
\mbox{Hen2-113}; (2) the observed spectra have a S/N ratio of 920 per
pixel   in the ordinary as well as in the extra-ordinary channel; (3)
statistical noise is the only source of errors; (4) both emission and
absorption lines are included in the analysis and `feel' the same Zeeman 
splitting. Under these assumptions, the value obtained for
$\left<B_{\rm z}\right>$ with FORS\,2 via our method of
analysis has a theoretical $1\,\sigma$ error of $\approx 35$\,Gauss. 
Note that this error is independent of the magnetic field strength 
assumed in modeling the simulated Stokes $V$ spectrum (here 
$\left<B_{\rm z}\right>=0$). 

This exercise suggests that the small formal $1\,\sigma$ error obtained
from the actual measurements of \mbox{Hen2-113} with methods R1 and RM in 
the range $\sigma_B = 18$ to $30$\,G is not unrealistic, though it may 
be somewhat underestimated. Note, however, that a number of lines seen in
the observation are missing in the synthetic spectrum.

\end{appendix}

\end{document}